\documentclass[aps,prb,superscriptaddressm,twocolumn]{revtex4}

\usepackage{amsmath,amssymb,bm}
\usepackage{graphicx}
\usepackage{epstopdf}
\usepackage{latexsym}
\usepackage{subfigure}
\usepackage{color}
\usepackage{wasysym}

\newcommand{\be}{\begin{equation}}
\newcommand{\ee}{\end{equation}}
\newcommand{\bea}{\begin{eqnarray}}
\newcommand{\eea}{\end{eqnarray}}

\newcommand{\bw}{\begin{widetext}}
\newcommand{\ew}{\end{widetext}}

\begin{document}
\title{ Metal-insulator transition in a two-band model for the perovskite nickelates}
\author{SungBin Lee}
\author{Ru Chen}
\affiliation{Department of Physics, University of California, Santa Barbara, CA-93106-9530}
\author{Leon Balents}
\affiliation{Kavli Institute for Theoretical Physics, University of
  California, Santa Barbara, CA, 93106-9530}

\date{\today}

\begin{abstract}
  Motivated by recent Fermi surface and transport measurements on
  LaNiO$_3$, we study the Mott Metal-Insulator transitions of
  perovskite nickelates, with the chemical formula RNiO$_3$, where R
  is a rare-earth ion. We introduce and study a minimal two-band
  model, which takes into account only the e$_g$ bands. In the weak to
  intermediate correlation limit, a Hartree-Fock analysis predicts
  charge and spin order consistent with experiments on R=Pr, Nd,
  driven by Fermi surface nesting.  It also produces an interesting
  semi-metallic electronic state in the model when an ideal cubic
  structure is assumed.  We also study the model in the strong
  interaction limit, and find that the charge and magnetic order
  observed in experiment exist only in the presence of very large
  Hund's coupling, suggesting that additional physics is required to
  explain the properties of the more insulating nickelates,
  R=Eu,Lu,Y. Next, we extend our analysis to slabs of finite
  thickness. In ultra-thin slabs, quantum confinement effects
  substantially change the nesting properties and the magnetic
  ordering of the bulk, driving the material to exhibit highly
  anisotropic transport properties. However, pure confinement alone
  does not significantly enhance insulating behavior.  Based on these
  results, we discuss the importance of various physical effects, and
  propose some experiments.
\end{abstract}

\maketitle

\section{Introduction} 
\label{sec:intro}

The Mott Metal-Insulator Transition (MIT) is a central subject in the
physics of correlated electron phenomena and transition metal
oxides.\cite{RevModPhys.70.1039}\ The perovskite nickelates, RNiO$_3$,
where R is a rare earth atom, constitute one of the canonical families
of materials exhibiting such an MIT.  One of the most interesting
features of the nickelates is the charge and spin ordering in the
insulating state, which is relatively complex yet in the ground state is
robust across the entire family.\cite{garcía1992sudden, PhysRevB.50.978, PhysRevB.57.456,
  PhysRevB.64.144417, rodriguez1998neutron,fernandez2001magnetic}   The explanation of this ordering is
still in many ways controversial.  While the MIT in bulk nickelates is
an old subject, the topic has been reinvigorated recently by attempts to
grow thin film heterostructures and observe unique quantum confinement
effects.\cite{chaloupka2008orbital,hansmann2009turning,son:062114,dan:opt-cond-LNO,gray:_insul_state_ultrat_epitax_lanio,kaiser:_suppr_fermi_lanio_srtio,PhysRevB.83.161102,liu2010strain,stewart2011optical,chakhalian2010origin}  In this paper, we revisit the problem of the MIT and ordering
in the nickelates, both in bulk and in heterostructures, from a very
simple theoretical viewpoint.  

We begin by summarizing some salient features of the nickelates.  First,
as the rare-earth ionic radius decreases, the MIT temperature
increases. Starting from R=La which is metallic at all temperature,
R=Pr, Nd have finite MIT temperatures T$_ {\rm MIT} = 120 $K, $180$K
respectively (R=Eu has the highest MIT temperature T$_{\text{\tiny MIT}}
= 480 $K) and finally R=Lu is insulating at all temperatures.  This
trend is understood due to the increasing distortions introduced in the
smaller rare earth materials, which increase the Ni-O-Ni bond angle and
hence reduce the bandwidth.  In the materials R=La, Pr, Nd, the
electrons can therefore be understood as more itinerant and bandlike,
while they are increasingly ``Mottlike'' for the smaller rare earths.

Second, at low temperature, all the nickelates display a magnetic
ordering pattern with an ``up-up-down-down'' spin configuration which
quadruples the unit cell relative to the ideal cubic
structure.\cite{garcía1992sudden, PhysRevB.50.978, PhysRevB.57.456,
  PhysRevB.64.144417, rodriguez1998neutron,fernandez2001magnetic} This
pattern coexists with a ``rock salt'' type charge order -- what is
actually observed is expansion or contraction of the oxygen octahedra --
which alternates between cubic sites.  Such charge order must indeed
always be present for this magnetic state, on symmetry grounds, and can
therefore be considered to this extent as a secondary order
parameter.\cite{lee2011landau} Interestingly, for the more metallic
nickelates both charge and spin order appear simultaneously, consistent
with this view, while for the more insulating nickelates, R=Eu, Ho, the
charge ordering occurs independently in an intermediate temperature
insulating phase without magnetism.\cite{rodriguez1998neutron,
  fernandez2001magnetic,girardot2008raman}

A variety of microscopic physical mechanisms have been proposed for the
nickelates.  A na\"ive view of the material would be to consider the
nickel d electrons only, occupying the nominal Ni$^{3+}$ valence state
which would place one electron in the e$_g$ doublet, which is degenerate
with cubic symmetry.  Early studies attributed the complex spin pattern
to orbital ordering, perhaps induced by Jahn-Teller or orthorhombic
distortions that split the e$_g$ degeneracy.  However, no Jahn-Teller
distortion was observed, and it was later suggested that orbital
degeneracy is removed by a separation of charge into Ni$^{2+}$ and
Ni$^{4+}$ states (an extreme view of the charge order), which have no
orbital degeneracy.\cite{scagnoli2006role}  This was attributed to strong Hund's rule exchange
on the nickel ion,\cite{mazin2007charge} but phonons may also be
involved.  However, the observed and robust magnetic ordering is not so
natural in this picture.  Another question mark is raised by
spectroscopic measurements, which seem to observe a significant of
Ni$^{2+}$ occupation,
suggesting that a model with holes on the oxygens may be more
appropriate.\cite{mizokawa2000spin}\ In this paper, we reconsider the
mechanisms for spin and charge ordering in the nickelates, and
specifically highlight the distinctions between an itinerant and
localized picture of the electrons.  Our main conclusion is that, at
least for the R=La, Pr, Nd materials where the MIT transition temperature is low or
zero, and in which a broad metallic regime is observed, the itinerant
picture is more appropriate.  We summarize the main content of the paper
below.  

The contrasts between the aforementioned models are really sharp only
deep in the Mott limit, in which orbital degeneracy, ionic charge, and
Hund's rule versus superexchange are clearly defined and distinct.  In
an itinerant picture, the precise atomic content of the bands is not in
itself important, but rather the physics should be constituted from a
model of the dispersion of the states near the Fermi energy and the
interactions amongst these same states.  In this view, the observed
ordering may be considered as spin and charge density waves (SDWs and
CDWs), and are tied to the Fermi surface structure.  Recent soft X-ray
photoemission\cite{eguchi2009fermi} indeed observed large flat regions
of Fermi surface in LaNiO$_3$, which appear favorable for a
nesting-based spin density wave instability.  

Specifically, in Sec.~\ref{sec:tight-binding} we introduce a minimal
two-band model for the electronic states near the Fermi energy in the
nickelates.  While it is easiest to motivate such a model from the
na\"ive view of Ni$^{3+}$ valence states -- which is questionable, as
noted above -- it can be considered just as the simplest
phenomenological tight binding Hamiltonian which can produce electronic
bands with the appropriate symmetry, in agreement with LDA
calculations\cite{hamada1993electronic}.  Within this model, the crucial
parameter controlling the shape of the bands is the ratio of the second
neighbor to first-neighbor d-d hopping.  With a small and reasonable
ratio, the large closed Fermi surface observed in experiments and LDA
calculations is reproduced.\cite{eguchi2009fermi,hamada1993electronic}
In addition, the same fermiology reasonably explains the resistivity,
Hall effect and thermopower measurements on LaNiO$_3$\cite{son:062114},
as well as the main features of the optical conductivity below
2eV.\cite{dan:opt-cond-LNO} We study the effect of interactions in this
model by a simple random phase approximation (RPA) criterion for the spin
density wave instability, and by more detailed Hartree-Fock
calculations, in Sec.~\ref{sec:hartree-fock}.  These mean-field type approaches are, we believe,
reasonably appropriate for the itinerant limit.  Interestingly, we find
that the same hopping ratio which reproduces the experimental Fermi
surface also turns achieves nearly optimal nesting, which further
supports the itinerant view.  The Hartree-Fock calculations then predict
the phase diagram as a function of spin-independent and spin-dependent
interactions, which we include microscopically by Hubbard $U$ and Hund's
rule $J_H$ couplings in the tight-binding model.

We find that the Hartree-Fock calculations produce two possible
explanations for the observed spin and charge ordering in the more
itinerant nickelates.  Theoretically, these two scenarios can be best
understood by considering a hypothetical ideal cubic sample (the real
materials undergoing MITs are orthorhombic even in the metallic state).
In such a sample, we obtain two distinct insulating ground states,
characterized by ``site centered'' and ``bond centered'' SDWs.  If it
occured within an otherwise cubic sample, the bond centered SDW would
have equal magnitude of moments on all sites, and would not induce
charge ordering.  In the site centered SDW, charge ordering is present,
and there would be a vanishing moment on one rock salt sublattice.  In
real orthorhombic samples, the bond centered SDW will be driven
off-center, and charge order is induced.  The latter off-center SDW
appears most consistent with experiment.  It is also the most favorable
SDW state in the Hartree-Fock calculations, and dominates in the regime
of relatively small $J_H$ coupling.  

For completeness, in Sec.~\ref{sec:strong-coupl-limit} we study the
two-band model in the strong coupling limit, in which $U$ and/or $J_H$
are much larger than the bandwidth.  In this limit, we find that an
insulating state with charge order consistent with experiment can be
obtained, but only for very large Hund's exchange, $J_H/U > 4$.  The
magnetic order is found to be either ferromagnetic or of the site
centered SDW type.  While the latter is quite close to what is
observed in experiment, it does not appear fully consistent, and
moreover the requirement of such large $J_H$ to stabilize a charge
ordered state seems to reaffirm the unphysical nature of this limit.

After this detour to strong coupling, we return to the reasonably
successful model and Hartree-Fock approach, and apply it to finite
thickness slabs in Sec.~\ref{sec:hetero}.  This provides a minimal and
highly idealized model for a nickelate film.  We find that quantum
confinement leads to substantial changes of the nesting properties of
ultra-thin slabs.  The predicted consequences are modified magnetic
ordering compared to bulk and and highly anisotropic transport
properties.  One result we {\sl do not} find from this calculation is a
substantial enhancement of the Mott insulating state in films of just a
few monolayers, a phenomena for which there is gathering experimental
evidence.\cite{son:062114,boris2011dimensionality,gray:_insul_state_ultrat_epitax_lanio,kaiser:_suppr_fermi_lanio_srtio,PhysRevB.83.161102}
We take this as evidence that the putative Mott insulating state in
ultrathin LNO films is driven not only by confinement but by additional
interface-sensitive effects.  

Finally, we conclude in Sec.~\ref{sec:opt-cond} with a discussion of
experiments, models, and some open issues.  In particular we discuss the
role of oxygen 2p orbitals, and a possible physical mechanism behind the
insulating state.  We also describe some experimental probes of the Mott
transition which may help to distinguish different mechanisms.

\section{Two-Band Model and Nesting Properties}
\label{sec:tight-binding}

The simplest tight-binding model for the nickelates is constructed based
on the na\"ive Ni$^{3+}$ valence.  In this ionic configuration, the only
partially occupied orbitals are the two members of the e$_g$ doublet,
containing one electron.  We consider the hopping through the
neighboring oxygen $p$ states ($\sigma$-bonding) as dominant, and treat
it as virtual.  This leads to strongly directional hopping, described as
\be H_{\rm tb} = - \sum_{ij} t_{ij}^{ab} c_{i a \sigma}^{\dagger} c_{j b
  \sigma}
\label{eq:tight-binding}
\ee where $i,j$ are site indices, $a,b=1,2$ are orbital indices for
$2z^2-x^2-y^2$ and $x^2-y^2$ respectively and $\sigma= \uparrow,
\downarrow$ is the spin index.   Comparison with LDA band calculation and
with the experimentally measured Fermi surface indicates that the
nearest-neighbor hopping $t$ and the next-nearest-neighbor hopping $t'$
with $\sigma$-type bonding is the most dominant. In detail, $t_{i, i \pm
  \hat{\mu} }^{ab} = t \phi_{\mu}^a \phi_{\mu}^b$ and $t_{i,i \pm
  \hat{\mu} \pm \hat{\nu} } = t' ( \phi_\mu^a \phi_\nu^b + \phi_\mu^b
\phi_\nu^a ) $ where $\phi_x = (- \frac{1}{2}, \frac{ \sqrt{3}}{2}),
\phi_y = (- \frac{1}{2} , - \frac{ \sqrt{3}}{2} ), \phi_z = (1,0)$ are
the orbital wavefunctions for the $2 x^2-y^2-z^2, 2 y^2-x^2-z^2$ and
$2z^2-x^2-y^2$ $\sigma$-bonding orbitals along the three axes. We
estimated $t'/t \approx 0.05$ by fitting our tight-binding model with
LDA band calculation, while the best fits to experimentally measured
Fermi surface gives $t'/t \approx
0.15$.\cite{hamada1993electronic,eguchi2009fermi} 

The range $0.05 \leq t'/t \leq 0.2$ reasonably explains the observation
of an hole-like Hall coefficient but an electron-like thermopower in
LaNiO$_3$.\cite{son:062114, xu1993resisitivity} This apparently
contradictory behavior of the Hall conductivity and thermopower arises
from the mixed electron and hole character on the e$_g$ Fermi
surface.   The model also predicts inter-band optical spectral weight in
 reasonable  correspondence with experiment at low energy (less than
 2eV).\cite{dan:opt-cond-LNO} 

\begin{figure}
  \centering
\includegraphics[width=3.0in]{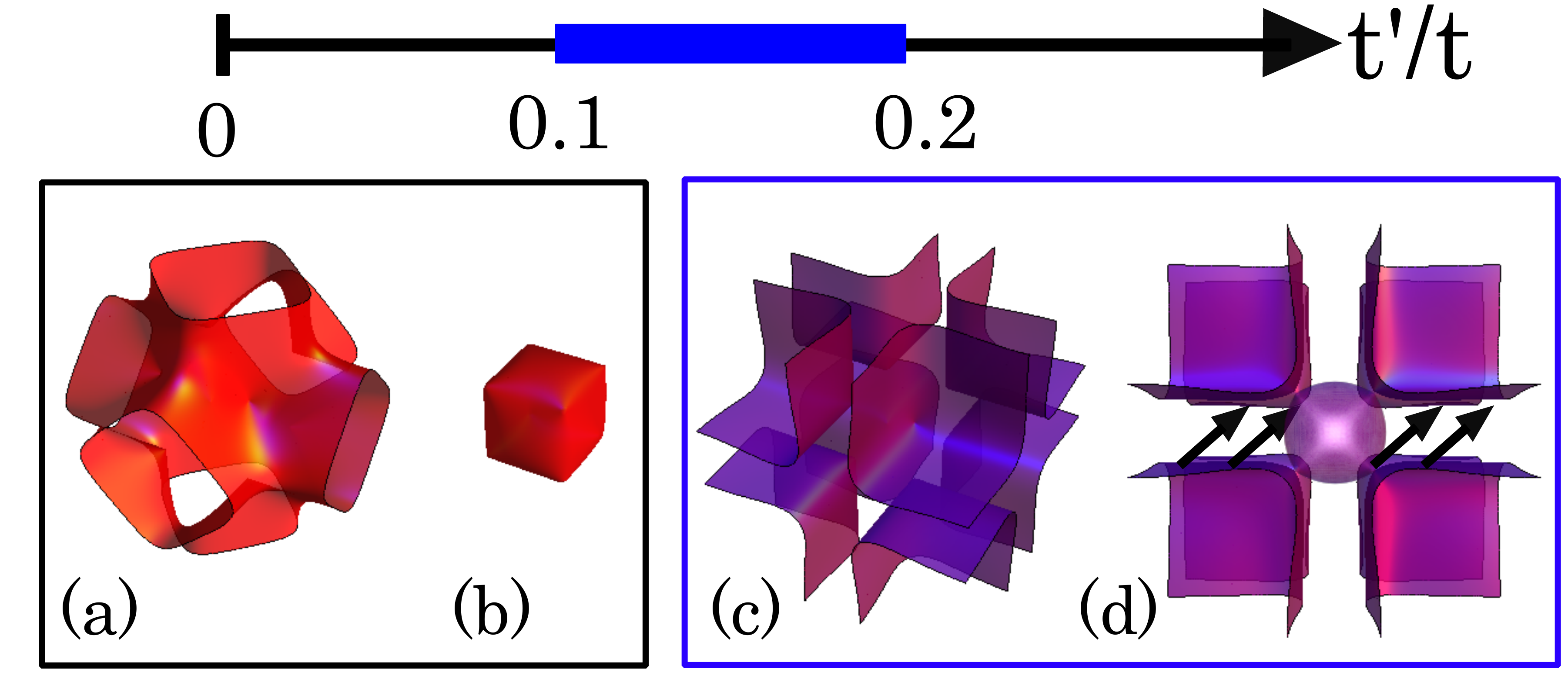}
\caption{Fermi surfaces for the tight-binding model.  In  (a) and (b),
  we show the conduction and valence band Fermi surfaces, respectively,
  for $t'/t = 0$.   For larger $t'/t$, the conduction band Fermi
  surfaces become large and hole-like, as shown in (c) and (d) for
  $t'/t=0.15$.  The approximate nesting in the latter case is indicated
  schematically in (d).   }
  \label{fig:FS}
\end{figure}

\begin{figure}
  \centering
\includegraphics[width=3.5in]{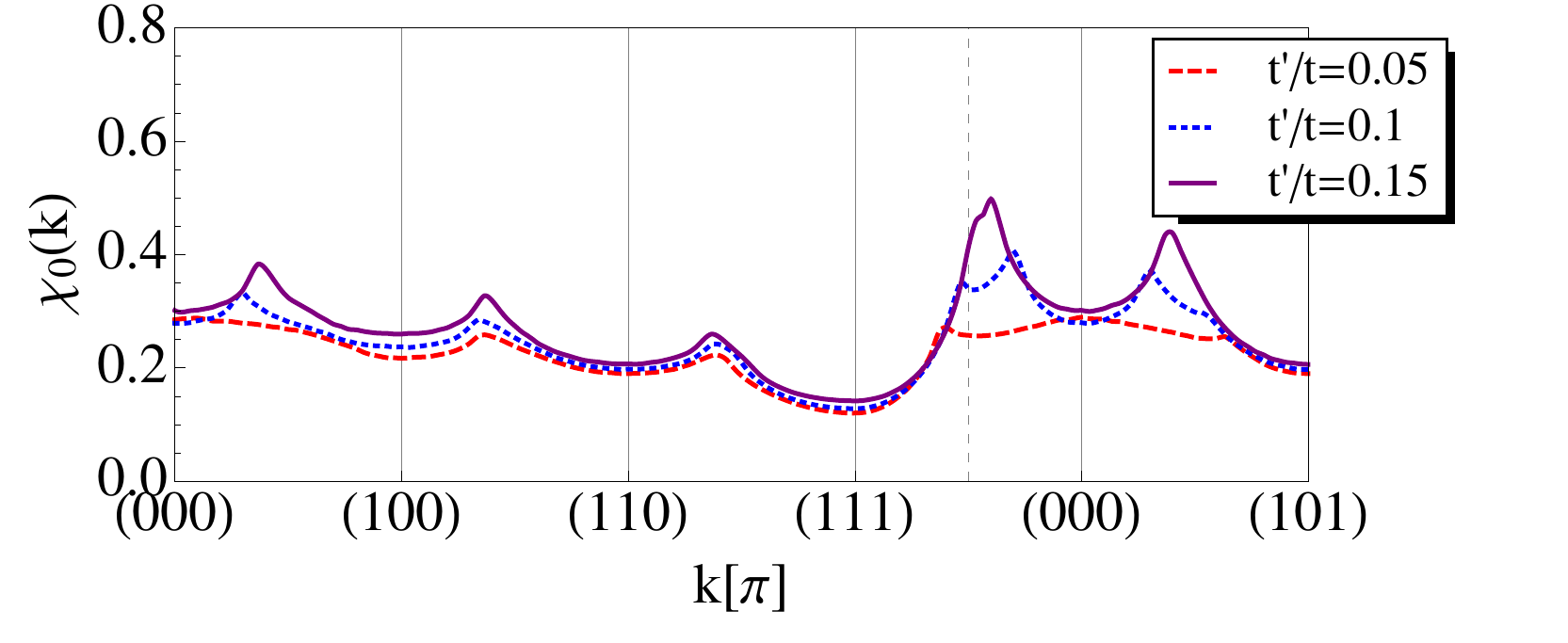}
\caption{Zero frequency spin susceptibility for the tight-binding
  Hamiltonian for $t'/t=0.05,0.1,0.15$, as a function of momentum ${\bf
    k}$ in the cubic Brillouin zone.  Note that for the best nested
  situation, $t'/t=0.15$, the susceptibility is sharply peaked close to
  the wavevector $2\pi(\frac{1}{4},\frac{1}{4},\frac{1}{4})$. }
  \label{fig:chi}
\end{figure}

We now examine the Fermi surface in more detail in search of nesting
tendencies.  Fig.\ref{fig:FS} shows representative Fermi surfaces
obtained from the tight-binding model as a function of the ratio
$t'/t$. With increasing $t'/t$ from $0$ to $0.15$, the topology of the
large Fermi surface is changing as seen in Fig.\ref{fig:FS}(a), (c) and
(d). In the absence of next-nearest-neighbor hopping $t'/t =0$, the
conduction band Fermi surface has an open topology as seen in
Fig.\ref{fig:FS} (a). With increasing $t'/t$, this Fermi surface becomes
closed, comprising a large ``pocket'' centered at the zone corner.  In
the intermediate range (especially $0.1 \leq t'/t \leq 0.2$), the pocket
resembles a cube, as seen in Fig.\ref{fig:FS}(c) and (d) ((d) shows both
valence band and conduction band Fermi surfaces). Contrary to the
conduction band Fermi surface, the valence band Fermi surface retains
its spherical topology for all $t'/t$ ( see Fig.\ref{fig:FS}(b)). The
experimental Fermi surface of LaNiO$_3$ observed by Eguchi at al
strongly resembles Fig.\ref{fig:FS}(d).\cite{eguchi2009fermi}  

The presence of large flat regions leads to nesting, and a tendency for
CDW and/or SDW instabilities.\cite{harrison1980electronic}   A simple
understanding of the effect of nesting is obtained from the Random Phase
Approximation (RPA), in which the effect of interactions on the spin
susceptibility is approximated by
\begin{equation}
  \label{eq:1}
  \chi(\omega,{\bf k}) = \frac{\chi_0(\omega,{\bf k})}{1- U
  \chi_0(\omega,{\bf k})},
\end{equation}
where $ \chi_0(\omega,{\bf k})$ is the non-interacting spin
susceptibility, and we took for simplicity a spin and
momentum-independent interaction $U$.  An instability is signalled by a
divergence of $\chi(0,{\bf k})$, which occurs on increasing $U$ when the
denominator in Eq.~\eqref{eq:1} vanishes.  This occurs for the ${\bf k}$
which maximizes $\chi_0({\bf k}) \equiv \chi_0(0,{\bf k})$, which determines the wavevector of
the spin ordering.  In the case of perfect nesting, $E_{\bf q} = E_{\bf
  q+k}$ for every ${\bf q}$ on Fermi surface with the nesting vector
${\bf k}$, and the non-interacting susceptibility is itself divergent at
this nesting wavevector, indicating an instability for arbitrarily small
$U$.  Although this is not true in general due to imperfect nesting, the
flatness of the Fermi surface greatly strengths the tendency to
instability.  

To check this directly, we calculate the zero frequency spin
susceptibility, which in general in the Matsubara formulation is given
by
\bw
  \bea \chi_0 ( i \omega_n , {\bf k} )
&=& \langle S^z_{\bf k} S^z_{\bf -k} \rangle  \\
&=& \frac{1}{2} \int \frac{d^3q}{ (2 \pi)^3} \frac{1}{\beta}
\sum_{\Omega_n} \text{Tr} [ G_0 (i \Omega_n, {\bf q}) G_0( i(\Omega_n +
\omega_n) ,{\bf  q}+{\bf k}) ],
 \label{eq:chi}
 \eea
 \ew
 where the free electron Green's function is defined as $G_0 ( i
 \omega_n , q) = \langle c^\dagger_{\bf q} c_{\bf q} \rangle = ( i
 \omega_n - E_q ) ^{-1}$.  More details are explained in
 Appendix.\ref{app:chi_0}. Fig.\ref{fig:chi} shows the calculated zero
 frequency spin susceptibility $\chi_0 ( {\bf k} )$ as a function of
 ${\bf k}$ for different ratios of $t'/t$. As expected, the spin
 susceptibility is sharply peaked at a particular certain wave vector in
 the physical range of $t'/t$. Specifically, $\chi_0 ({\bf k})$ for
 $t'/t = 0.15$ shows the highest peak at ${\bf k}={\bf Q}_n=2 \pi
 (\frac{1}{4}, \frac{1}{4}, \frac{1}{4})$ which defines the nesting
 vector. Note that this is precisely the magnetic ordering wavevector
 (in the cubic convention) observed in the insulating low temperature
 phase of the nickelates.  Estimating the instabilty from
 Eq.~\eqref{eq:1}, we obtain $U_c \approx 1 / \chi_0 ( {\bf Q}_n)
 \approx 2 $ (see Fig.\ref{fig:chi}).

\section{Hartree-Fock theory}
\label{sec:hartree-fock}

\subsection{Restricted Hartree-Fock Method}

Having established the nesting wavevector, we proceed to a (restricted)
Hartree-Fock treatment of the ordering and MIT.  We include interactions
in the two-band model via an on-site Coulomb term $U$ and Hund's
coupling $J_H$, defined from $H  =  H_{\rm tb} + H_{\rm int}$,
\begin{eqnarray}
H_{\rm int} & = & U \sum_{i} n_i^2 - J_H \sum_{i} {\bf S}_i^2, \label{eq:6} 
\end{eqnarray} where $n_i = \sum_{a \alpha} n_{i
  a \alpha}$ and $ {\bf S}_i = \sum_{a \alpha \beta} c_{i a \alpha
}^{\dagger} \frac{{\boldsymbol \sigma}_{\alpha \beta}}{2} c_{i a \beta}
$.  As discussed earlier, what is important here, because of the nesting
physics, is the interaction between states near the Fermi surface.  As
such, the $U$ and $J_H$ terms may be thought of as simply a convenient
parametrization of the spin-independent and spin-dependent parts of
these interactions, rather than literally in terms of atomic Coulomb and
Hund's rule terms.  

\begin{figure}
  \centering
\includegraphics[width=3.5in]{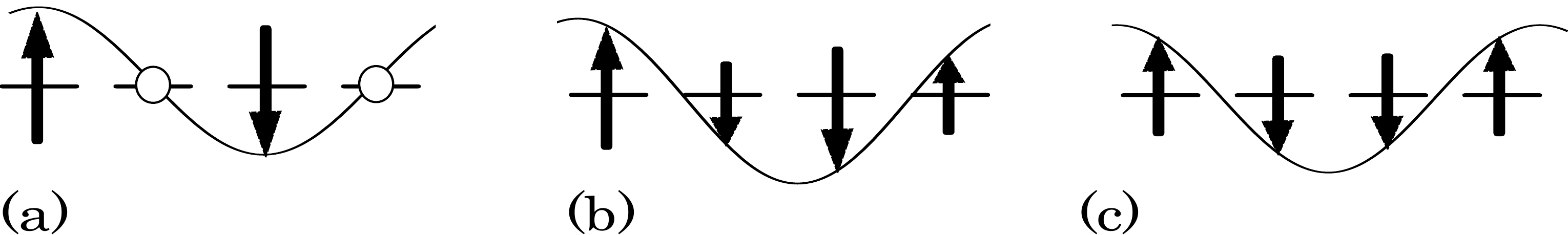}
\caption{Spin configurations depending on the phase of $\psi$, $\theta$, along $\hat{x}$ axis. (a) shows ``site-centered" spin ordering for $\theta=0$, (b) for intermediate $\theta = \pi/8$ and (c) is ``bond-centered" ordering for $\theta = \pi/4$}
  \label{fig:spin-conf}
\end{figure}

To treat the problem in Hartree-Fock, we define a variational
wavefunction as the ground state of a fiducial mean-field Hamiltonian,
which has the form of a non-interacting two-band hopping model {\sl
  plus} linear ``potentials'' arising from coupling to SDW and CDW order
parameters.  Experimental results predominantly favor collinear magnetic
ordering, of the form 
\be
\langle {\bf S}_i \rangle \propto  {\bf h}_i = \hat{z} {\rm Re} [ {\boldsymbol \psi} e^{i {\bf Q}_n \cdot {\bf r}_i } ]
\label{eq:spin-conf}
\ee
with complex variable ${\psi} \equiv |\psi | e^{i \theta}$.  
Fig.\ref{fig:spin-conf} shows different spin configurations which depend
on the phase of $\theta$.  For instance, $\theta = 0$ corresponds to
``site-centered" spin ordering in which the spin pattern is
``up-zero-down-zero'' moving along a cubic axis, while $\theta=\pi/4$
gives ``bond-centered" ordering, and an ``up-up-down-down'' pattern. In
the intermediate regime $0 < \theta <\pi/4$, the order is ``off-center''
as shown in Fig.\ref{fig:spin-conf}(b). 

As already discussed above and in Ref.\onlinecite{lee2011landau}, a CDW order
parameter will be induced with ${\bf Q}_{\rm cdw} = 2 {\bf Q}_{n} = \pi
(1,1,1) $ as observed in experiment. This charge ordering is commonly
known as ``rock-salt" ordering and implies the electron density at site
$i$ is represented as
 \be
\label{eq:CO}
\langle n_i \rangle \propto \rho_i = (-1)^{x_i + y_i + z_i} \Phi,
\ee
where $\Phi$ is an Ising-type order parameter for the charge ordering.  

The full mean-field Hamiltonian from which the Hartree-Fock variational
ground state is constructed then takes the form
\bea
H_{\rm var} &=& \tilde{H}_{\rm tb} + H_{\rm dw} ,\\
H_{\rm dw} &=&  - \sum_i {\bf h}_i \cdot {\bf S}_i -
\sum_i \rho_i n_i .
\label{eq:Hvar}
\eea The local exchange field ${\bf h}_i$ and the charge ordering
$\rho_i$ couple to the spin operator ${\bf S}_i$ and the electron number
operator ${\bf n}_i$ respectively. Note that we allow additional freedom
in the variational state by letting the hopping parameters renormalize.
That is 
\begin{equation}
  \label{eq:2}
  \tilde{H}_{\rm tb} = H_{\rm tb}[t\rightarrow \tilde{t}, t'\rightarrow
  \tilde{t}']. 
\end{equation}

The restricted Hartree Fock calculation proceeds by finding the ground
state of $H_{\rm var}$:
\begin{equation}
  \label{eq:3}
  H_{\rm
  var} | \Psi_0 \rangle = E_0 | \Psi_0 \rangle,
\end{equation}
with the constraint of quarter-filling, i.e. one electron per site,
$\sum_i n_i = N$, where $N$ is the number of sites.  The Hartree-Fock ground state
$|\Psi_0\rangle$ is then a function of four
dimensionless parameters: $ \tilde{t}'/\tilde{t}, | \psi| / \tilde{t},
\Phi / \tilde{t} $ and $\theta$. 
 For each set of parameters, we calculate the variational energy 
 \begin{equation}
   \label{eq:4}
   E_{\rm
  HF} = \langle \Psi_0 | H | \Psi_0 \rangle,
 \end{equation}
which is then minimized over the dimensionless parameters, for fixed
physical parameters $t,t' ,U$ and $J_H$.  

To find $|\Psi_0\rangle$ in practice, we work in the reduce Brillouin
zone (BZ) determined by the four site magnetic unit cell.  We thereby
end up with instead of two bands 8 magnetic ones, constructed from the
different pieces of the original BZ folded into the magnetic one, 
\begin{equation}
  \label{eq:5}
c_{n a \alpha} ({\bf k}) = c_{a
  \alpha}({\bf k} + n {\bf Q}_{\rm sdw} ),
\end{equation}
with $n=0,1,2,3$ (for four magnetic sublattices), where $a$ is for two
e$_g$ orbitals and $\alpha$ is for spin $\uparrow \downarrow$.   In this
basis,
\be {\tilde H}_{\rm tb} = \sum_{\bf k}^{'} \sum_n {\tilde H}_{ab}(
{\bf k}+n {\bf Q}_{\rm sdw} ) c_{n a \alpha}^{\dagger} ({\bf k}) c_{ n b
  \alpha} ({\bf k})
\label{eq:tildeH-tb-8}
\ee
The prime on ${\bf k}$ sum means the sum over the reduced BZ.  In the same way, the density wave Hamiltonian in ${\bf k}$ space is represented as,
\bea
H_{\rm dw} &=& \sum_{\bf k}^{'} \sum_n \frac{ \alpha \psi}{4} c_{n+1 a
  \alpha}^{\dagger} ({\bf k}) c_{n a \alpha} ({\bf k}) + {\rm h.c.} \nonumber \\
   & &  \hspace{1cm}+ \Phi c_{n+2 a \alpha}^{\dagger} ({\bf k}) c_{n a \alpha} ({\bf k})
\label{eq:H-dw-8}
\eea
We then find the single-particle eigenstates by diagonalizing the $
8 \times 8$ matrix of the variational Hamiltonian Eq.~\eqref{eq:Hvar}, and
construct $|\Psi_0\rangle$ by filling the states up to the Fermi energy,
determined by the requirement of $1/4$ filling.   It is then
straightforward to express $E_{\rm HF}$ in terms of the single-particle
states and occupation numbers, and perform the minimization procedure (see
Appendix.\ref{app:detail-HF} for more details).  

\subsection{Hartree-Fock Phase Diagram}

\subsubsection{Two SDW states}
\label{sec:two-sdw-states}

\begin{figure}
  \centering
\includegraphics[width=3.5in]{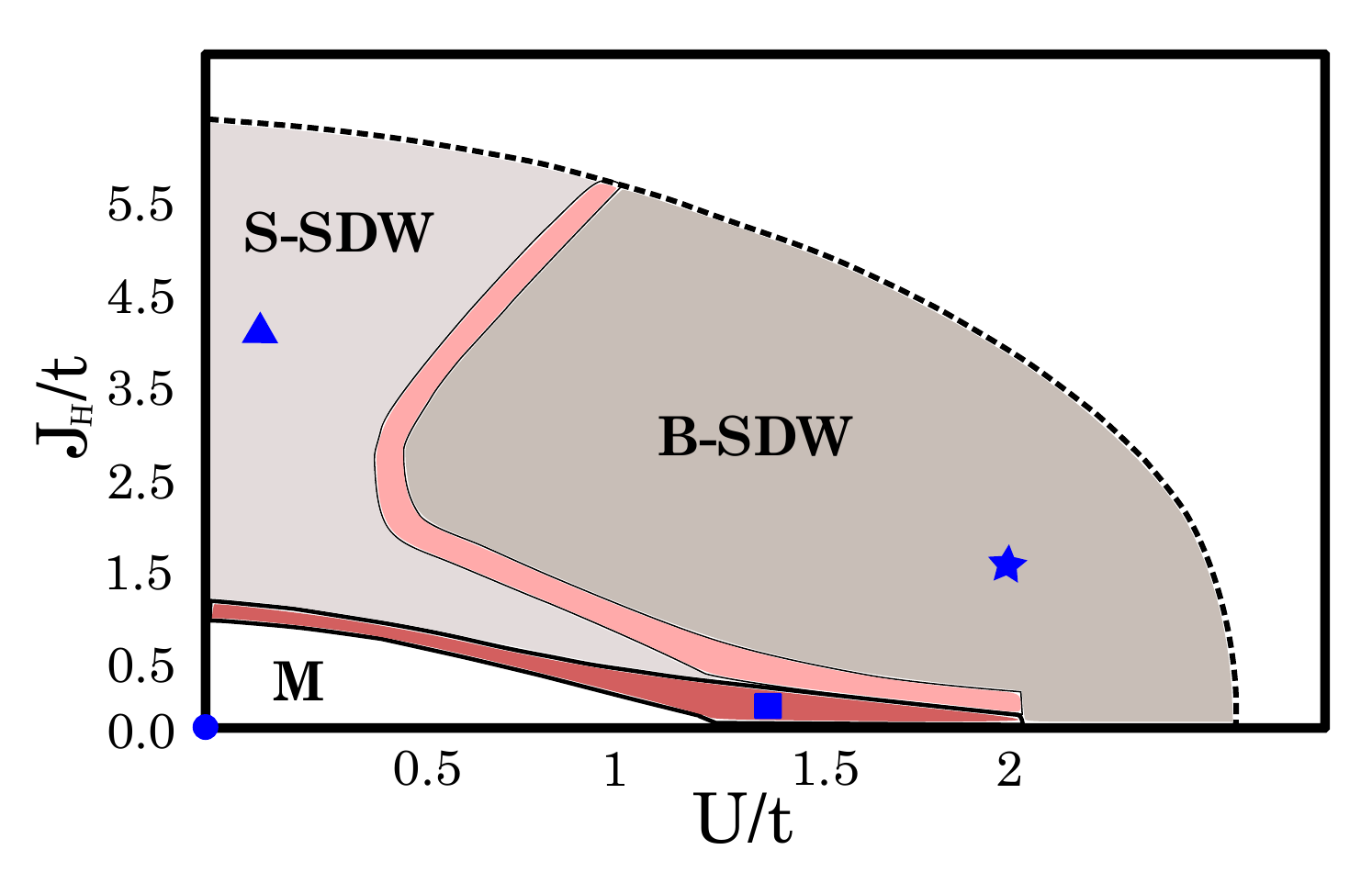}
\caption{MIT bulk phase diagram for the model with cubic symmetry as a
  function of $U/t$ and $J_H/t$, with $t'/t=0.15$.  Here $U$ is the
  on-site Coulomb interaction, $J_H$ is the Hund's coupling and $t$ is
  the nearest-neighbor hopping magnitude.  The main phases which appear are
  a paramagnetic metallic state (\textbf{M}), metallic SDW (dark pink
  region close to \textbf{M}), insulating site-centered SDW
  (\textbf{S-SDW}), and semi-metallic bond-centered SDW
  (\textbf{B-SDW}).  In between the
  \textbf{S-SDW} and \text{B-SDW} state, one observes an off-center
  SDW phase with $0 < \theta< \pi/4$ (pink region). The blue
  colored shapes show the points for which the optical
  conductivity is plotted in Sec.\ref{sec:opt-cond}}
 \label{fig:mit-pd}
\end{figure}

The resulting Hartree-Fock phase diagram for a typical situation with
$t'/t$=0.15 is shown in Fig.~\ref{fig:mit-pd}.  We observe a metallic
regime at small $U$ and $J_H$, and two main ordered phases with stronger
interactions.  For large $J_H$, site-centered SDW ordering with
$\theta=0$ occurs, concurrent with strong charge order, generating
an insulating state.  This is natural because the large $J_H$ favors
pairing of electrons into spin $S=1$ moments, requiring neighboring
empty sites.  More mathematically, such Hartree-Fock states minimize the
Hund's term.  For large $U$, the bond-centered SDW with $\theta=\pi/4$
occurs instead.  This is again natural because the $U$ term prefers
uniform charge density, and with $\theta=\pi/4$ in the cubic system
(which we discuss here) no CDW order occurs.

\subsubsection{semi-metallic B-SDW}
\label{sec:semi-metallic-b}

Somewhat surprisingly, the bond-centered SDW state remains semi-metallic
even at relatively large $U$ within the Hartree-Fock approximation.
Indeed, examination shows that the density of states is almost linearly
vanishing approaching the Fermi energy in this region, with a small
non-zero value at $E_F$, which decreases with increasing $U$.  This
unusual behavior arises from the specific ``up-up-down-down" magnetic
ordering in this phase.  To understand it, recall that the cubic
lattice, viewed from the $[ 111 ]$ direction, forms stacks of triangular
lattice layers.  In the limit of strong bond-centered ordering, the
spins on each triangular plane are fully polarized. Moreover, electrons of one spin polarization
are confined to a pair of parallel $[ 111]$ planes, which together forms
a honeycomb lattice when connected by the dominant nearest-neighbor
hopping $t$.  Thus in the limit of large $U/t$ in the bond-centered SDW
state, the appropriate tight-binding model is that of doubly
degenerate e$_g$ orbitals on honeycomb lattice.  \bea H_{\hexagon} =
-\sum_{ij} t^{ab}_{ij}c^{\dagger}_{i a} c_{j b}
\label{eq:H_hexagon}
\eea This model has four orbitals per unit cell due to the doubly
degenerate e$_g$ orbitals and the bipartite honeycomb lattice.
Fig.\ref{fig:band-111}(a),(b),(d) and (e) shows the dispersion and the
DOS of this tight-binding model for the cases $t'/t = 0$ and
$0.15$. Without second-nearest-neighbor hopping $t'/t =0$, the result
contains two bands which are identical to those of the canonical
nearest-neighbor tight-binding model for graphene, possessing two Dirac
cones with linear dispersion at Fermi level.   The similarity with
graphene has led to the suggestion that such systems might be used to
engineer a topological insulator.\cite{xiao2011interface}  With increasing $t'/t$, the
DOS saturates at a small non-zero value approaching the Fermi
level. This is because finite $t'/t$ introduces both
second-nearest-neighbor hopping and, more importantly, coupling between
the honeycomb bilayers.  The latter expands the Dirac points into small
electron and hole pockets, in a similar manner as inter-layer coupling
does in graphite.  

\subsubsection{effects of orthorhombicity}
\label{sec:effects-orth}

As discussed in Ref.\onlinecite{lee2011landau}, the bond-centered ordering
in the large $U$ region is actually unstable to orthorhombicity
(GdFeO$_3$ distortion) , which is present in all the nickelates save
LaNiO$_3$.  This is expected on symmetry grounds to drive the SDW
off-center.  The off-centering in turn induces charge order.  Thus at
the symmetry level, when orthorhombicity is taken into account, the
large $U$ region is completely consistent with experiment.

What of the metallicity in this region?  In the graphene-like honeycomb
bilayer, the Dirac point degeneracy is protected, as it is in graphene,
by inversion symmetry.  Inversion is indeed preserved by the
bond-centered SDW in the ideal cubic system.  It is, however, violated
when both the SDW and orthorhombic distortion are present.  Hence, we
expect that orthorhombicity not only affects the centering of the SDW,
it also tends to open a gap in the electronic density of states,
converting the semi-metal to a true insulator.

\begin{figure}
 \centering
\includegraphics[width=5.in]{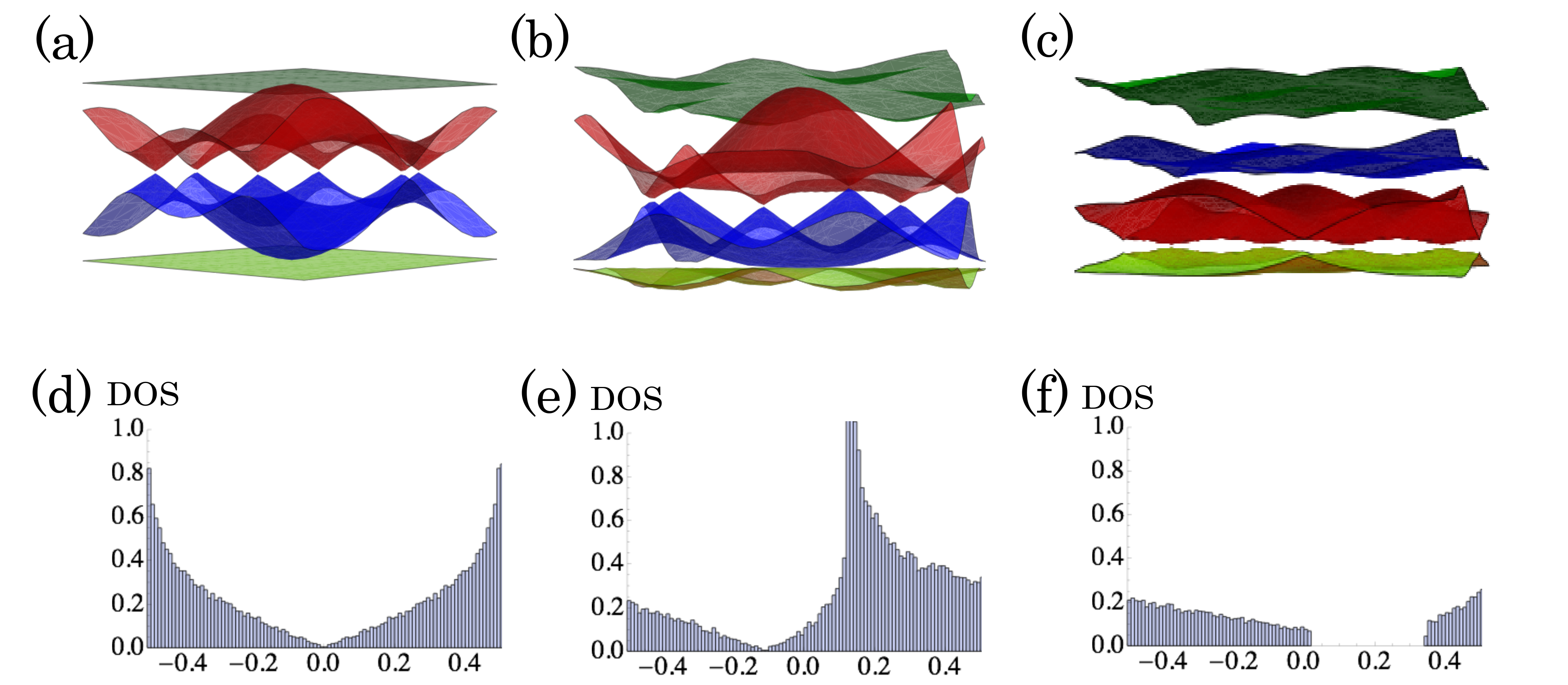}
\caption{Dispersion (panels (a),(b) and (c)) and density of states (DOS, panels
  (d),(e),(f)) for the e$_g$ tight-binding model on honeycomb
  lattice. In (a) and (d), $t=1$ and $ t' =0$, we observe Dirac points
  with clear linear dispersion and corresponding linear DOS.  The
  Dirac cone is stable to small $t'=0.15$ as shown in (b) and (e).  In
  (c) and (f), an orbital splitting induced by the orthorhombic
  distortion of the lattice is included,  with $t'=0.15$ and ${\bf
    D}=1.5/ \sqrt{3} (1,1,1)$. An induced gap is clearly seen. }
  \label{fig:band-111}
\end{figure}

\begin{figure}
  \centering
\includegraphics[width=2.5in]{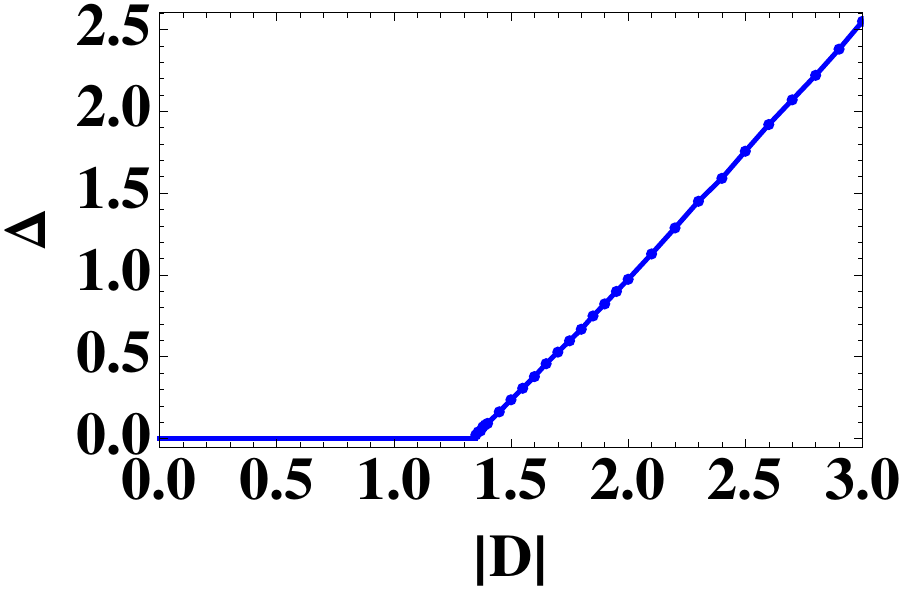}
\caption{Plot of the single-particle gap $\Delta$ versus the orbital
  field $|D|$, for the honeycomb model with nearest-neighbor $t=1$ and
  $t' = 0.15$.  Here we have arbitrarily taken the orbital field
  of the form ${\bf D} = (D,D,D)/\sqrt{3}$. }
  \label{fig:D-gap}
\end{figure}

We now study this microscopically.  A leading effect of the orthorhombic
distortion is expected to be a crystal field splitting of the e$_g$
orbitals at each Ni site. Therefore, we add the on-site orbital
splitting term \bea H_{\text{ortho} } = \sum_i {\bf D}_i \cdot
c^{\dagger}_{i a} \tau_{ab} c_{i b}
\label{eq:H_ortho}
\eea
Here we have suppressed the (diagonal) spin indices, and introduced Pauli
matrices ${\tau}$ in the orbital space.  Using the symmetries of the
Pbnm space group of the orthorhombic structure, we find (see
Appendix.\ref{app:ortho}) that the
``orbital fields'' ${\bf D}_i$ are all expressible in terms of a single
vector ${\bf D}$  :
\bea
{\bf D}_i = ( (-1)^{x_i + y_i} D^x  , (-1)^{x_i + y_i} D^{y}  , D^{z} )
\label{eq:D_i}
\eea For simplicity, we consider this term in the effective honeycomb
lattice model, Eq.\ref{eq:H_hexagon}, relevant for the large $U$
case. Fig.\ref{fig:band-111} shows how the the DOS changes in the
presence of an orthorhombic distortion.  A gap indeed opens for
sufficiently large ${\bf D}$, as plotted in Fig.~\ref{fig:D-gap}.  

\subsubsection{Limitations of the restricted HF theory}
\label{sec:limit-restr-hf}

Because we consider a {\sl restricted} Hartree-Fock ansatz, some lower
energy states that do not fit this ansatz may be missed in
Fig.\ref{fig:mit-pd}.  For example, near the onset of SDW order, at
relatively weak interactions, there is the possibility of an {\sl
  incommensurate} SDW.  This may be expected since the best nesting
vector determined by the maximum of the susceptibility is not {\sl
  exactly} at the commensurate value, but rather at $ {\bf Q} \approx
0.4 \pi (111)$ (see Fig.\ref{fig:chi}).  Generally, commensurate states
are preferred at strong coupling, and if incommensurate phases exist,
they would be expected to change to the commensurate ones with
increasing interaction strength, via a commensurate-incommensurate
transition.\cite{chaikin2000principles}

We have also neglected the possibility of spontaneous orbital
ordering, which could occur in the cubic model at large $U$.
Indeed, orbital degeneracy is crucial to the semi-metallicity
found in the B-CDW phase, as we have seen above via the introduction of
orthorhombicity.  Spontaneous orbital splittings (ordering) provide a
mechanism for the cubic model to achieve a truly insulating state, which
it must at sufficiently large $U$.  However, we argue that the absence
of any observed orbital ordering or Jahn-Teller distortion is evidence
that this physics is not relevant for the nickelates.

\

\bw

\ew

\section{Strong coupling limit}
\label{sec:strong-coupl-limit}

The Hartree-Fock approach of the previous section is reasonable for weak
to intermediate strength interactions, which we believe is most relevant 
for the more itinerant nickelates with R=Pr,Nd.   For completeness, in
this section we study the complementary limit of strong interactions,
$U/t, J_H/t \gg 1$.  Here the two-band model is suspect, so the connection
to experiment is less clear.  However, we can at least qualitatively
attempt to address the question of the interplay of charge and spin
order in the strong coupling regime.  Specifically, note that
in the more insulating nickelates, with R=Eu,Ho,\cite{medarde1992rnio_}
charge ordering appears {\sl first} upon lowering temperature from the
paramagnetic metallic state, with magnetism occuring only at lower
temperature.  Thus it seems that in these materials there is a separation of
scales, with the primary mechanism for the MIT being charge ordering,
and magnetism being secondary.  In this section, we will see that this
is indeed the case in one regime of the strong coupling limit of the two
band model.  The specific parameters of this region do not, however,
seem very physical, supporting the idea that in the more insulating
nickelates a description beyond the two band model is needed.  

The strong coupling limit may be considered an expansion in the hopping
$t,t'$ about the limit $t=t'=0$.  In the extreme limit, the behavior is
determined entirely by the ``atomic'' Hamiltonian $H_{\rm int}$ in
Eq. \eqref{eq:6}, which can be solved independently at each site,
subject to the constraint of proper total electron occupation (quarter
filling).  There are two regimes, determined by the parameter
$\alpha=U/J_H$.  For $\alpha>1/4$, the atomic ground state is one with one
electron per site.  In this regime every site is equivalent, and has
four states available to it, due to the spin and orbital degeneracy.
Further perturbation in $t,t'$ will therefore result in a spin-orbital
Hamiltonian of the Kugel-Khomskii type. 

The other regime occurs when
$\alpha<1/4$, and in this case the electrons prefer to segregate into
two sets of sites with equal numbers in each: doubly occupied sites with
total spin $S=1$, and empty sites.  The ground state energy in this
regime is $E_0=-(1-2\alpha)NJ_H$, where $N$ is the number of sites.  Here
there are two sorts of degeneracies.  First, for $t=t'=0$ the {\sl
  location} of the paired sites is undetermined, so there is a
degeneracy of $N!/[(N/2)!]^2$ associated with the different possible
location of the pairs.  In addition, for each of the paired sites, there
are 3 spin states available.

In the remainder of this section, we will focus on this latter regime.
Physically, we may consider the paired sites as {\sl bosons} with spin
$S=1$.  By introducing hopping perturbatively, we may introduce hopping
and interactions between the bosons.  In the perturbative treatment,
we will, in addition to $t/U, t/J_H \ll 1$, further assume $t'/t \ll
1$, which simplifies the algebra considerably.  Below, we argue that the leading
effects of hopping, at $O(t^2)$, induce {\sl charge ordering} of the
bosons, reducing the problem to an effective spin $S=1$ model.  The spin
degeneracy of the bosons is split only at the next non-trivial order,
$O(t^4)$.  This qualitatively agrees with the separation of scales
observed between charge and spin order in the nickelates.

\subsection{$O(t^2/J_H)$: charge ordering}
\label{subsec:elec-loc}

We first consider the effective Hamiltonian for the system at the
leading non-vanishing order in perturbation theory, which is second
order in hopping, for the case of $\alpha<1/4$.  To formulate the
perturbation theory , we treat the Hund's and Coulomb part as the
unperturbed Hamiltonian, $\mathcal{H}_0= H_{\rm int}$, and the hopping
as the perturbation, $\mathcal{H}_1 = H_0$.  We denote the projection
operator onto the ground state manifold of $\mathcal{H}_0$ at quarter
filling by ${\mathcal P}$.  If $|\Psi\rangle$ is an exact eigenfunction
of the system with energy $E$, then its projection into the ground state
subspace, $|\Psi_0\rangle = \mathcal{P} |\Psi\rangle$ satisfies
\be
  \label{eq:60}
  \left[E_0 + \mathcal{P} \mathcal{H}_1 \frac{1}{1-\mathcal{R}\mathcal{Q}\mathcal{H}_1}\mathcal{R}\mathcal{H}_1 \right]|\Psi_0\rangle = E|\Psi_0\rangle, 
\ee
 where $\mathcal{R}= (\mathcal{H}_0-E)^{-1}$
is the resolvent and $\mathcal{Q}=1-\mathcal{P}$ .  Eq.~\eqref{eq:60} is an implicit non-linear eigenvalue problem and we will only evaluate it perturbatively in $\mathcal{H}_1$, then it becomes
\begin{eqnarray}
  \label{eq:61}
 \mathcal{H}_{\rm eff}|\Psi_0\rangle  &\equiv&  (E-E_0)|\Psi_0\rangle \nonumber \\
 &\approx& \mathcal{P}\mathcal{H}_1 \sum_{n=0}^3 (-1)^n \left(\mathcal{R} \mathcal{Q} \mathcal{H}_1\right)^n |\Psi_0\rangle,
\end{eqnarray}
where to this order of accuracy, we can safely approximate
$\mathcal{R}\approx  (\mathcal{H}_0-E_0)^{-1}$. 

The second order term in degenerate perturbation theory corresponds to $n=1$ in Eq.~\eqref{eq:61}, in which electrons make two
consecutive virtual hopping transitions. The terms for three different types of hops can be combined (see Appendix.\ref{app:DPT} for more details),
up to an additive constant, into
\begin{eqnarray}
  \label{eq:63}
  \mathcal{H}_{\rm eff}^{(1-3)} & = & \sum_{\langle ij\rangle} \big[ \frac{4t^2}{J_H}\frac{1}{1-4\alpha}+\nonumber \\
  &&\frac{2t^2}{J_H}\frac{1}{5+4\alpha} 
 \left (\vec{S}_i \cdot \vec{S}_j -1\right) \big]N_i N_j.
\end{eqnarray}  

Eq.~\eqref{eq:63} gives the effective Hamiltonian at leading order for
$\alpha<1/4$.  To solve it, we note that $N_i$ commutes with
$\mathcal{H}_{\rm eff}^{(1-3)} $ and is thus a good quantum number {\sl at every site}.
We then can easily see that the charge ordered states with $N_i=0,2$
on the two rock salt fcc sublattices saturate a lower bound on the energy,
of $\mathcal{H}_{\rm eff}^{(1-3)} =0$.  This follows because, since
the eigenvalues of $\vec{S}_i \cdot \vec{S}_j$ are bounded by -2,
hence the effective boson-boson repulsion (the term in the square
brackets in Eq.~\eqref{eq:63}) obeys
\be
  \label{eq:64}
  V_{\rm eff}=\left\langle \frac{4t^2}{J_H}\frac{1}{1-4\alpha}+\frac{2t^2}{J_H}\frac{1}{5+4\alpha} \left (\vec{S}_i \cdot \vec{S}_j -1\right) \right\rangle>0.  
\ee
Thus, regardless of the specific spin states of the boson pairs, their
nearest-neighbor interaction is always repulsive for
$0\leq\alpha<\frac{1}{4}$.  The lower bound and hence charge order
in the ground state follows. 

\subsection{Magnetic interactions}
\label{subsec:spin-state-boson}

Notably, although the effective interaction $V_{\rm eff} \sim t^2/J_H$
determines the charge order in the ground state (and defines the
energy scale separating it from uniform states), the spin degrees of
freedom on the doubly occupied sites remain undetermined at leading
order.  The spin physics is dictated by subdominant terms.  Thus the
appearance of charge order at a higher temperature than magnetism is a
feature of this limit of the two-band model.  Let us now consider the
magnetic interactions in more detail.

First we focus on the spin exchange between nearest-neighbor sites on
the fcc sublattice, i.e. second nearest-neighbor sites on the original
cubic lattice.  There are three lowest orders that we will consider:
$O(t^4/J_H^3)$, $O(t^2t'/J_H^2)$, $O(t'^2/J_H)$. Although the effects of the
$t'$ hopping is a relatively small correction to the dominant $t$
hopping, in the strong $J$ limit, it is not negligible because it can
contribute at second and third order to the exchange between spins.
Formally all these terms are on an equal footing if we take $t' \sim
t^2/J_H$. We combine the contributions from different orders together (see Appendix.\ref{app:DPT} for more details). The total spin exchange between nearest-neighbor sites on the fcc sublattice is 
\bw
\be
\label{eq:J1}
 J_1 = -\frac{t^4}{J_H^3}\frac{1}{(1-4\alpha)^2}\big[\frac{8(5+4\alpha)}{(1-4\alpha)(5-4\alpha)}-\frac{5}{5+4\alpha}-\frac{1}{1-4\alpha}\big] +\frac{t^2t'}{J_H^2}\frac{1}{1-4\alpha}\left[\frac{10}{5+4\alpha}+\frac{5}{1-4\alpha}\right]+ \frac{t'^2}{J_H}\frac{5}{5+4\alpha}.
\ee
\ew

 Next we focus on the spin exchange between next nearest-neighbor sites.  Calculation then shows 
 \be
  \label{eq:70}
 J_2 =\frac{t^4}{J_H^3}\frac{1}{(1-4\alpha)^2}\left[\frac{16}{5-4\alpha}+\frac{8}{5+4\alpha}\right] 
\ee

\begin{figure}
  \centering
\includegraphics[width=3.in]{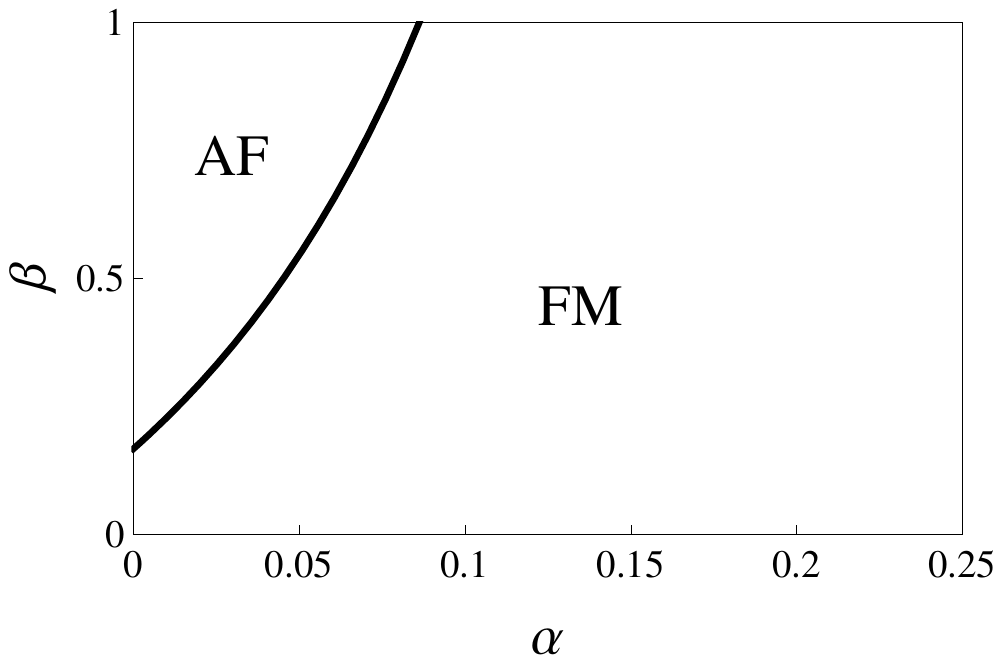}
\caption{Phase diagram of the classical ground state as a function of two dimensionless parameters $\alpha\equiv U/J_H$ and $\beta\equiv t'J_H/t^2$. }
  \label{fig:spin-state}
\end{figure}

From the expression of $J_1$ and $J_2$, we obtain that if both $t'$ and $\alpha$ are resonably small, there's ferromagnetic interactions between nearest-neighbors and antiferromagnetic interactions between second nearest-neighbors on the fcc lattice. Case (1) for $O(t^4/J_H^3)$ term is the dominant term for the ferromagnetic interaction. The negative sign for that case can be understood as arising
due to the Hund's rule coupling on the intermediate $k$ site, which
prefers the two transferred virtual electrons to be in a triplet
state.  For the $J_2$ exchange, however, because $i,j,k$ are all along
a single cubic axis, only one orbital can hop.  For this reason, in the first hopping procedure (which is dominant) that contribute to $J_2$ it is impossible to obtain a triplet intermediate state, since two electrons in a single orbital must form an antisymmetric singlet.  This explains the antiferromagnetic sign of this exchange.

Let us see what magnetic structure is expected from this exchange
Hamiltonian.  Since the fcc lattice is a Bravais lattice, we can use
the Luttinger-Tisza method to find the classical ground states.  We
simply Fourier transform the exchange couplings to obtain the energy
of spiral states with wavevector ${\bm k}$.  One finds
\begin{eqnarray}
  \label{eq:68}
  E_k & = & -6 J_2 + 4 J_2 \sum_{\mu=1}^3 x_\mu^2 + 4 J_1
  \sum_{\mu>\nu} x_\mu x_\nu,
\end{eqnarray}
where $x_\mu = \cos k_\mu$.  Since this energy is quadratic in the
$x_\mu$, we can consider it as a quadratic form.  The eigenvalues of
the form are $8(J_1+J_2)$ and $8J_2-4J_1$ (the latter is twofold
degenerate).  It is therefore positive definite if $J_2>{\rm Max}
(-J_1,J_1/2)$.  When this is satisfied, the minimum energy states are
those with $x_\mu=0$, i.e. $k_\mu = \pm \pi/2$.  These are exactly the
magnetic states observed experimentally. The phase diagram in Fig.\ref{fig:spin-state} shows the classical magnetic ground state for different values of $\alpha$ and $\beta\equiv t'J_H/t^2$. We note that
if $t'=0$ ($\beta=0$) and $\alpha=0$, the ground state appears to be ferromagnetic.  When $t'$($\beta$) is included, the ferromagnetic $J_1$
interaction is decreased, and the antiferromagnetic state will be stabilized. For the region $\alpha<0.1$, the magnetic ground state is antiferromagnetic when $t'\sim t^2/J_H$. It is
remarkable that one can obtain in this way the same magnetically
ordered state as found from the itinerant nesting picture.

\subsection{ Comparison with weak coupling limit}
\label{subsec:comparison}

According to the perturbation theory of the large Hund's coupling,
charge order first appears at $O(t^2/J_H)$ and then magnetic ordering
occurs due to perturbation at $O(t^4/J_H^3)$,$O(t'^2/J_H)$ and
$O(t^2t'/J_H^3)$ . Since the magentic ordering arises from a temperature
scale smaller than the charge ordering phase, this agrees with the
experimentally observed intermediate charge ordering phase without
magnetism. On the other hand, in the weak coupling limit, the charge
ordering is always slaved to the primary magnetic
ordering.\cite{lee2011landau}


\section{Confinement effects in thin films}
\label{sec:hetero}

The success of the Hartree-Fock theory in reasonably predicting the
charge and spin ordering in the more itinerant nickelates undergoing a
MIT suggests that the approach may also be profitably applied to films.
Recently, various growth issues have been overcome leading to epitaxial
films of good quality on several substrates with layer by layer control.
One may expect that the MIT and related charge and spin ordering can be
strongly modified in thin films, due to both distortions (dependent on
details of the substrate and growth conditions), effects of changes in
chemistry at interfaces, and to quantum confinement effects.  Because of
the difficulty of controlling the former two effects (which in any case
are better studied by first principles methods), we focus here entirely
on the latter, and consider in this section the simplest possible model
of a finite thickness film.  That is, we simply take the bulk
tight-binding Hamiltonian and apply it to a finite thickness slab
consisting of $L$ unit cells in the confined direction, with effectively
``vacuum'' outside the slab, i.e. open boundary conditions.  Given the
importance of Fermi surface shape in determining the nesting properties,
we expect that quantum confinement alone can significantly modify the
MIT properties and the ordering in the insulating state.

\subsection{Single Layer, $L=1$}

\begin{figure}
  \centering
\includegraphics[width=7in]{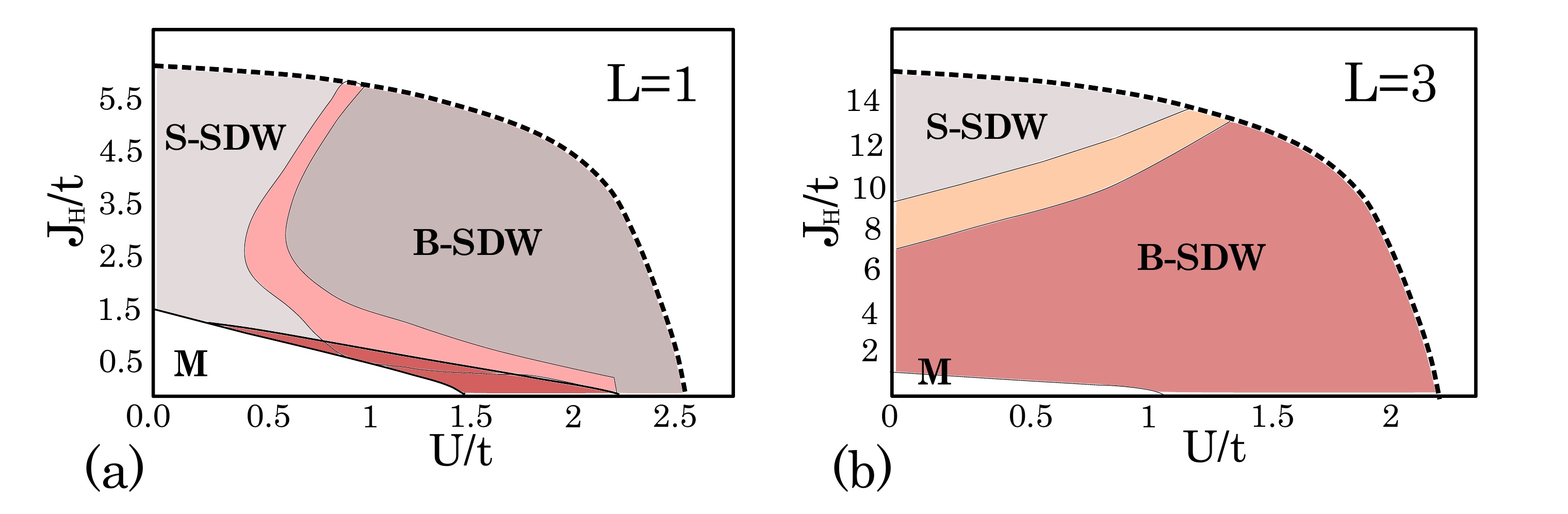}
\caption{Panel (a) shows the zero temperature phase diagram for a
  single layer $L=1$ with the nesting vector ${\bf Q}^{\rm
    2d}_{\text{sdw} } = 2\pi (\frac{1}{4} , \frac{1}{4} )$. As in the
  bulk case, the paramagnetic metallic phase (\textbf{M}) is stable
  for weak interactions.  The dark pink region indicates a metallic
  SDW state, and the light pink region indicates an insulating
  off-center SDW.  Panel (b) shows the phase diagram for three layers,
  $L=3$, with the nesting vector ${\bf Q} ^{L=3}_{\text{sdw}}= 2
  \pi(\frac{1}{4},0,0)$. This nesting vector leads to a metallic
  \textbf{B-SDW} phase, which persists even for large $U/t$. The yellow
  region between the \textbf{S-SDW} and \textbf{B-SDW} phases is a
  metallic off-center SDW.  }
  \label{fig:mit-pd-2d}
\end{figure}

First of all, we consider the extreme case of a single NiO$_2$ layer,
following the methods used for the bulk.  Here and throughout this
section, we will neglect the symmetry-lowering effects that must be
present in such a two dimensional structure, and in particular any
tetragonal orbital splitting which is likely to be the dominant effect
of this type.  With this proviso, the Fermi surface and nesting
properties are shown in Fig.\ref{fig:fin-lay}(a).  The two dimensional
Fermi surfaces show large flat regions similar to the bulk case. The
zero frequency spin susceptibility, $\chi_0^{\rm 2d} (k)$, is shown in
Fig.\ref{fig:fin-lay}(e) (see Appendix.\ref{app:chi_0}).  It is
sharply peaked at ${\bf Q}_{\rm sdw}^{\rm 2d} = 2\pi (\frac{1}{4},
\frac{1}{4})$.  Repeating the Hartree-Fock calculations for this case,
using this SDW vector, we obtain the phase diagram in
Fig.\ref{fig:mit-pd-2d}(a).  The results are quite similar to the bulk
case, except that the bond-centered SDW is insulating in this case, as
the honeycomb lattice structure does not arise for a single square
lattice layer.  Somewhat surprisingly, the location of the MIT
$(U/t)_c \approx 2 $ at $J_H/t=0$ remains almost unchanged from the bulk
case.  Na\"ively, one would expect a decrease in $(U/t)_c$ in 2d,
because the bandwidth is reduce by confinement.  We attribute the lack
of such a decrease to decreased nesting in the two dimensional case,
as can be seen by comparing Fig.\ref{fig:chi} and
Fig.\ref{fig:fin-lay}: the susceptibility has a higher peak in bulk
than in the single layer.  

\

\bw 

\ew

\subsection{Intermediate thickness films}
\label{sec:L3}

We now consider the intermediate cases with $L\geq 2$ NiO$_2$ layers
along the $\hat{z}$ direction.  In this case, the single-particle states
can be taken as standing waves in the vertical ($\hat{z}$) direction, with
$k_z = \pi l / ( L +1)$ where $l= 1,2, \cdots ,L$.  One obtains
correspondingly $2L$ subbands ($2$ arising from the orbital degeneracy),
each of which may have a Fermi surface. The calculated non-interacting
Fermi surface and spin susceptibility for several values of $L$ are
shown in Fig.\ref{fig:fin-lay} (see Appendix.\ref{app:chi_0} for more
details of the calculation of the spin susceptibility).   

From Fig.\ref{fig:fin-lay}(g), we see that the peak of the
susceptibility varies considerably and in a non-monotonic fashion with
$L$.  While the case $L=2$ (orange points in Fig.\ref{fig:fin-lay}(f))
is quite similar to the result for the single layer, $L=3,4$ are
considerably distinct.  For larger $L$, there is a slower variation of
behavior, and by increasing the thickness to $L=30$ (purple points in
Fig.\ref{fig:fin-lay}(f)), the bulk behavior (black line in
Fig.\ref{fig:fin-lay}(f)) is almost perfectly recovered.  Thus we expect
particularly distinct phase diagrams for the cases $L=3,4$, and focus on
these below.

\subsubsection{$L=3$}
\label{sec:l=3}

For $L=3$, one observes comparable peaks in the susceptibility at two wavevectors: 
${\bf Q} = \pi/2 (100)$ and ${\bf Q} = \pi/2
(110)$.  The former is quite distinct from the ordering in the single
layer and bulk cases.  To decide amongst the two possibilities, we
compared the variational energy in the Hartree-Fock approximation for
the two choices, and found that, over the full range of $U$ and $J$, the
total energy is lower for ${\bf Q} = \pi/2 (100)$.  Thus the model
predicts quite distinct ordering in the trilayer case.  

The full Hartree-Fock phase diagram, assuming this wavevector, is shown
in Fig.\ref{fig:mit-pd-2d}(b).  Details of the calculations for finite
$L$, which are somewhat complicated by the many subbands, are given in
App.\ref{app:detail-HF}.  Once again both site-centered and
bond-centered SDW states appear, but the site-centered SDW occurs here
only at very large values of the Hund's coupling, $J_H/t \gtrsim 10$,
making it probably entirely unphysical.  Another distinction from the
cases discussed previous is that the bond-centered SDW for $L=3$ appears
to be fully metallic.  This is because the SDW with wavevector ${\bf Q}
= \pi/2 (100)$ describes stripes of electrons with all spins parallel in
vertical stripes along the $y$ direction.  Thus the electrons are free
to hop in this direction -- actually they form ``ladders'' of two
parallel spin-aligned chains -- and one has a sort of
quasi-one-dimensional metallic state.  Instabilities of the
one-dimensional ladders would probably be expected beyond the
Hartree-Fock approximation, and could lead to further charge/spin/orbital
ordering and insulating behavior, but this is not within the scope of
our study.

\subsubsection{$L=4$}

\begin{figure}
  \centering
\includegraphics[width=7in]{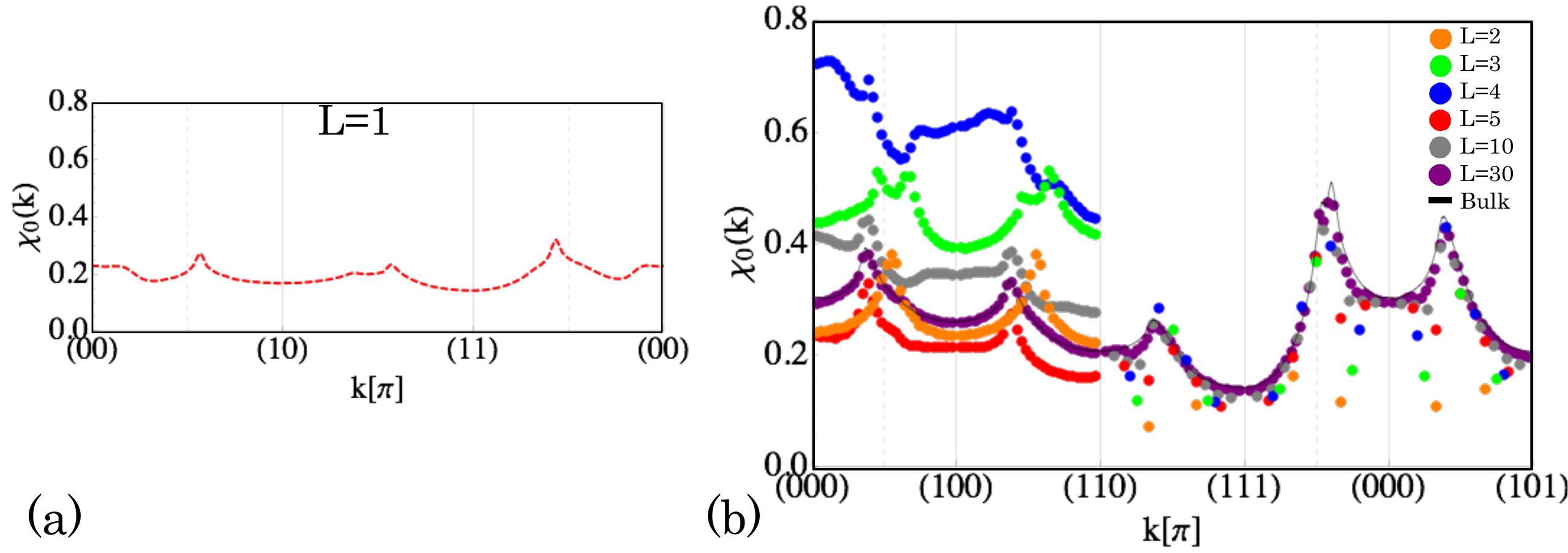}
\caption{The zero frequency spin susceptibility, $\chi_0({\bf k})$,
  for finite thickness slabs of $L$ layers, for the free electron
  tight-binding Hamiltonian with $t'/t=0.15$. Plot (a) shows the
  purely two-dimensional single layer case. Here $\chi_0({\bf k})$
  is sharply peaked at ${\bf k}={\bf Q}_{\text{sdw}}^{2d} = \pi/2
  (11)$. Plot (b) shows several cases with varying thickness with $2
  \leq L \leq 30$, compared with the bulk case $L=\infty$.  One sees
  that large $L=30$ (purple dots) agrees well with the bulk
  susceptibility (black solid line).   For smaller $L$, we see that
  the nesting properties change considerably.   This is especially
  pronounced for $L=3$ (green points), for which $\chi_0({\bf k})$ is sharply
  peaked at ${\bf k}={\bf Q} \approx \pi/2 (100)$, and for 
  $L=4$ (blue points), for which it is peaked at ${\bf k} \approx 0$.}
  \label{fig:fin-lay}
\end{figure}

\begin{figure}
  \centering
\includegraphics[width=6in]{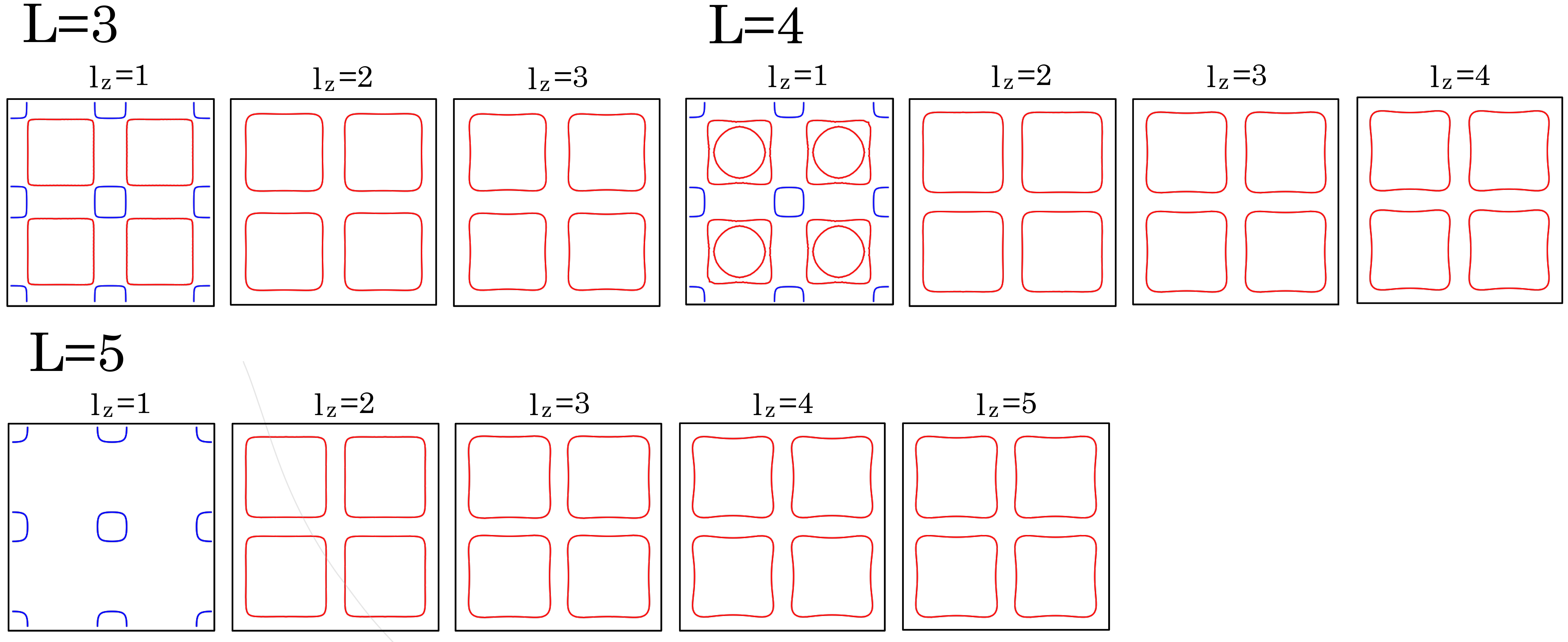}
\caption{The Fermi surfaces for different subbands in the $k_x$-$k_y$
  plane with discretized $k_z = \pi l_z / ( L+1)$ for the cases
  $L=3,4$ and $5$.  The blue (red) lines correspond to conduction
  (valence) subbands.  In the case $L=4$, we see two valence subband
  Fermi surfaces at $k_z = 0.2 \pi$ ($l_z=1$), which is responsible
  for an enhancement of the DOS at Fermi energy.  }
  \label{fig:subbands}
\end{figure}

One more noticeable feature in the spin susceptibility plotted
Fig.\ref{fig:fin-lay}(f), is the large ${\bf Q} \approx 0$ peak for
$L=4$ (see blue points).  The ${\bf Q}=0$ (uniform)
susceptibility is simply proportional to the density of states, which
is apparently enhanced for this film thickness.  The origin of this
enhancement is seen by inspecting
separately the Fermi surfaces associated with individual sub-bands
with discretized $k_z = \pi l_z / (L+1)$, shown in
Fig.\ref{fig:subbands}.  One sees that the $L=4$ case is unique in
having three distinct Fermi surfaces (two hole and one electron) for
the $l_z = 1$ sub-band.  Since the density of states is proportional
to the Fermi surface area, this explains the observed enhancement.
Some understanding of this is obtained by inspecting the bulk Fermi
surface, Fig.\ref{fig:FS}(d).  It contains a large hole-like surface
which has rather flat faces, parallel to [001] planes.  For the
specific case $L=4$ and $l_z=1$, the discretized $k_z = 0.2 \pi$ cuts
across this rather flat when $l_z =1$. As a result,
the flat face, leading to the multiple two-dimensional sub-band Fermi
surfaces.   This enhanced density of states could potentially lead to
ferromagnetism for this case, but since ferromagnetism is notoriously
over-estimated by the Hartree-Fock approximation, we do not pursue
this further here.

\bw

\ew

\section{Discussion}
\label{sec:opt-cond}

In the prior sections, we have studied a minimal two band model for
the perovskite nickelates, with a focus on the MIT and the spin and
charge ordering in the insulating state.  

\subsection{Do we need the oxygen orbitals?}
\label{sec:do-we-need}

In the minimal model used in this paper, we have eliminated the oxygen
orbitals to obtain an effective two orbital Hubbard model.  Several
papers in the literature, however, claim that the oxygen states are
crucial for the physics of the nickelates. Here we will discuss this
issue, and argue that the importance of explicit inclusion of the
oxygen states depends upon the questions being asked.

In general, in the Fermi liquid paradigm, which applies to weakly to moderately
correlated {\sl itinerant} systems, the behavior of the electrons is
dictated by the vicinity of the Fermi surface(s) only, and by the
effective interactions {\sl amongst these states near the Fermi
  surface}.  The great insight of Landau in developing Fermi liquid
theory was that the actual wavefunctions of these ``quasiparticle''
states are largely unimportant.  Thus when it applies, any model which
properly mimics the band dispersion near the Fermi surface (and its
symmetry), and which captures sufficiently the interactions amongst
the near-Fermi surface states, serves to correctly model the
electronic behavior.  It is well established now that LaNiO$_3$, the
metallic end-member of the RNiO$_3$ series, has a Fermi surface which
is obtained from the intersection of just two bands with the Fermi
energy.  These bands have $e_g$ character, which can be mimicked by
the minimal tight-binding model used in this paper.  Provided a band
picture of the important electronic states near $E_F$ is adequate,
this basis is sufficient to describe the nickelates.  The extent of
the microscopic oxygen versus nickel character of the states is
subsumed into the Bloch wavefunctions, {\sl which do not appear in the
  band Hamiltonian}, and to a lesser extent in the effective
interactions.  We conclude that for low to intermediate energy
properties for which the two-band description is adequate, explicit
treatment of the oxygen states is not important.

However, one may ask questions -- and conduct experiments -- for which
the oxygen states are obviously essential.  For instance, inelastic
x-ray scattering can measure the relative fraction of Ni$^{2+}$ and
Ni$^{3+}$ occupation of the Ni 3d states.  Estimates for NdNiO$_3$ is
that there is as much as 40\% Ni$^{2+}$.  By neutrality, the
Ni$^{2+}$ can only arise through the presence of holes on the oxygen
states.  This implies the Bloch wavefunction associated with the the
``oxygen bands'' and ``nickel bands'' have in fact considerably mixed
character.   However, this does not affect the reliability of the two
band model for the states near the Fermi energy.  Indeed, the
measurement of the Ni valence state is actually a measure of the
occupied states, and hence is really related to the character of the
filled valence band Bloch wavefunctions, not that of the near-Fermi
surface states.  Of course, by orthogonality, if the nominally oxygen states
have mixed character, so too must the nickel states.

Other high energy questions may be sensitive to the oxygen character.
For instance, let us consider the properties of an interface.  In
standard semiconductor systems, an interface can be understood through
band diagrams, which include only the energies of the bands, and not
their wavefunctions.  Thus when this approach applies, the oxygen
character is not important.   In fact, band diagrams rely upon a semiclassical
treatment which assumes that the electrostatic potential, carrier
density, etc. vary slowly with respect to the lattice spacing.  This
in turn is correct in semiconductors due to their small effective mass
and large dielectric constant.  There is no need for this to apply to
nickelate interfaces.

In fact, it would be natural to expect a change in the oxygen
character at an interface.\cite{han2010chemical} Consider an interface with a band
insulator such as LaAlO$_3$ (LAO), in which there are no 3d orbitals
near the Fermi energy.  By neutrality, in LAO the oxygen valence
should be ``exactly'' (or at least much more so than in the
nickelates) O$^{2-}$.  This implies that the Ni 3d orbitals in the
plane adjacent to the LAO are less able to hybridize with the
intervening oxygens, since these states are ``blocked''.  One can
consider a simple model in which this physics is accounted for by
ascribing an oxygen orbital energy $\epsilon_{p'}$ for the intervening
oxygens which is lower (so that here the electrons are more strongly
bound to their oxygen) than the energy $\epsilon_p$ for the same
orbitals inside the nickelate, i.e. $\epsilon_{p'} < \epsilon_p$.  The
larger energy separation $\epsilon_d - \epsilon_{p'}>
\epsilon_d-\epsilon_p$ for the interfacial states implies reduces
mixing of the nickel and oxygen states.  Thus we expect that {\sl the
  Ni$^{2+}$ character of the interfacial nickel ions should be
  reduced}.  As already remarked, this is a high energy property,
related to the occupied states.  However, the reduced mixing has
implications at low energy as well.  It implies reduced level
repulsion between the 3d (specifically the d$_{z^2}$) and 2p states,
so that the partially filled orbitals corresponding to the near Fermi
energy states should be lowered relative to bulk nickelates near the
interface.  That is, the conduction electrons feel an {\sl attraction}
to the d$_{z^2}$orbitals in the interfacial NiO$_2$ plane.  Note that,
although oxygen physics induces corrections to its Hamiltonian
parameters, the two-orbital model remains valid even for the
interface.

This physics may be relevant to recent experiments on LNO
heterostructures.  Several experiments have indicated the formation of
an insulating state for very thin LNO films with only a few unit
thickness.  This appears at odds with the calculations in
Sec.~\ref{sec:hetero}, which find that the metal-insulator transition
point is largely unchanged by confinement, even for very thin films.
This model, however, neglects the induced orbital potential at the
interface.  One would expect this orbital potential to partially
polarize the orbitals at the interface in favor of the d$_{z^2}$
states which conduct poorly in the xy plane. Moreover,
the shift of these orbitals renders inter-layer tunneling non-resonant,
which will further reduce the kinetic energy.  Thus it is natural to
expect the insulating state to be enhanced by this effect.  In
the future, we plan to investigate this in more detail by including
the interfacial orbital attraction explicitly in the Hartree-Fock
calculation.

\subsection{Strong versus intermediate correlation}
\label{sec:strong-vers-interm}

In this paper, we have contrasted the limits of weak to intermediate
correlation (and Hartree-Fock theory) and strong correlation (the
perturbative approach in Sec.~\ref{sec:strong-coupl-limit}).  It was
argued that the strong coupling limit seems not very realistic.
However, there are indications that something beyond the weak coupling
view is needed, at least for the more insulating nickelates, with
R=Lu,Ho,Y.  In these materials, the charge ordering and insulating
transition occurs above 500K but magnetism only sets in around 100K.
A factor of 5 or more discrepancy between these two scales is hard to
reconcile with a weak-coupling picture.  One type of strong-coupling
picture is discussed by Anisimov {\sl et al}\cite{PhysRevB.59.7901},
in which the nickel charge state is regarded as Ni$^{2+}$, which forms
an $S=1$ spin, while the mobile charge is actually in the form of
holes on the O sites.  The corresponding model would be a type of
underscreened Kondo lattice.  Charge ordering of the type seen in experiment is
certainly possible, and would be viewed as the formation of collective
Kondo singlets between two holes and a Ni$^{2+}$ spin on half the
lattice sites.\cite{sawatzky} To our
knowledge, whether this actually occurs for a Kondo model of this type
has not been established theoretically.  This is an interesting
problem for future study.  A likely issue with such a Kondo
description is that the band structure appears very different from the
bands with $e_g$ character predicted and observed in LaNiO$_3$.
Instead, the itinerant carriers must arise from oxygen bands, and it
is not clear why this should in any way mimic the $e_g$ structure.
But perhaps the bands in LuNiO$_3$ etc. {\sl are} radically different
from those in LaNiO$_3$.  If so, this should be testable
experimentally.

Some sort of intermediate coupling picture is also possible.  Indeed,
even if the most insulating materials are at strong coupling, and, as we
have suggested, PrNiO$_3$ and NdNiO$_3$ are better thought of in the SDW
(weak to intermediate coupling) limit, then there are compounds in
between.  Here presumably a full description with all the orbital
involved and charge fluctuations allowed in all orbitals is needed, and
there is little simplicity to be found.  Probably an approach which
combines elements of {\sl ab initio} theory and reasonable but ad-hoc
treatment of interaction physics such as DMFT is the best in this
regime.\cite{han2011dynamical} In this situation, it will unfortunately
probably be difficult to identify any single mechanism for charge
ordering.

In our opinion, it is likely that one physical effect we have not so
far discussed, the coupling to lattice phonons, is important.  The
Kondo singlet formation mentioned above would obviously benefit from a
contraction of the neighboring oxygens around the Ni$^{2+}$ spin in
question.  Indeed, it is this contraction which is actually observed
experimentally, rather than any real electric charge density.  The
same local phonon mode which would couple to the Kondo singlet would
also favor charge ordering in the intermediate coupling view.  It may
be that this electron-phonon interaction gives a reasonable mechanism
for the more insulating nickelates.

\subsection{Experimental signatures}
\label{sec:exper-sign}

It is desirable to understand how the different scenarios might be
distinguished experimentally.  We will focus here primarily on the
expected consequences in the itinerant regime, as the primary focus of
this work.  However, we briefly discuss expectations for the strong
coupling limits.  In the strong coupling pictures, we would presumably
expect the insulating states to have a full gap to electron and hole
quasiparticles.  Moreover, local $S=1$ moments would be well-formed on
half the Ni sites (forming an fcc sublattice), prior to ordering into
an antiferromagnetic ground state.  With these site-center local spins,
it seems difficult to imagine an antiferromagnetic state with the
symmetry of the bond-centered or off-center SDW, and we would expect a
site-centered SDW (antiferromagnetic) order.  This particular symmetry
could be distinguished by a careful determination of local moments at all
the nickel sites from neutron or NMR/$\mu$SR measurements.   

Turning now to the itinerant regime, we consider the experimental
consequences of the nesting scenario.  First we discuss the thermal
phase transition. In this limit, since the SDW drives the
charge order, the two types of order should set in simultaneously at a
single critical temperature.  In Ref.\onlinecite{lee2011landau}, it was
shown that this transition is theoretically expected to be first order
for several reasons.  These two observations are consistent with
experiment.  

More detailed comparison can be made with electronic structure.  We
discuss in particular the implications of the nesting scenario for dc
transport and optical measurements in the following.

\subsubsection{Transport anisotropy} 

Transport is an important probe of the electronic structure.  In the
nesting picture, the SDW order is directly and strongly coupled to the
quasiparticles, and hence should strongly influence the transport.
The most qualitative feature of this coupling is that the SDW order
imposes its lower lattice symmetry, and in particular, spatial
anisotropy, upon the quasiparticles.  In contrast, within the strong
coupling view, the charge ordering is dominant, and this charge
ordering itself is not anisotropic (it doubles the unit cell but is
compatible with cubic symmetry).  We therefore expect that, when the
nesting picture is valid, prominent transport anisotropy should be
observed to set in for $T< T_{\rm MIT}$.  

We focus first on the bulk case, for which the SDW wavevector ${\bf
  Q}_{\rm sdw}=2\pi(1/4,1/4,1/4)$ obviously breaks cubic symmetry.  As
discussed in Sec.~\ref{sec:semi-metallic-b}, the electronic structure
in the B-SDW phase is describable as a set of weakly coupled honeycomb
[111] bilayers, leading (neglecting orthorhombicity) to a
semi-metallic state.  Hence we expect the B-SDW ordering to be
accompanied by strong electrical anisotropy, with much larger
conductivity within the [111] plane than normal to it.  

\begin{figure}[hbtp]
\includegraphics[width=3in]{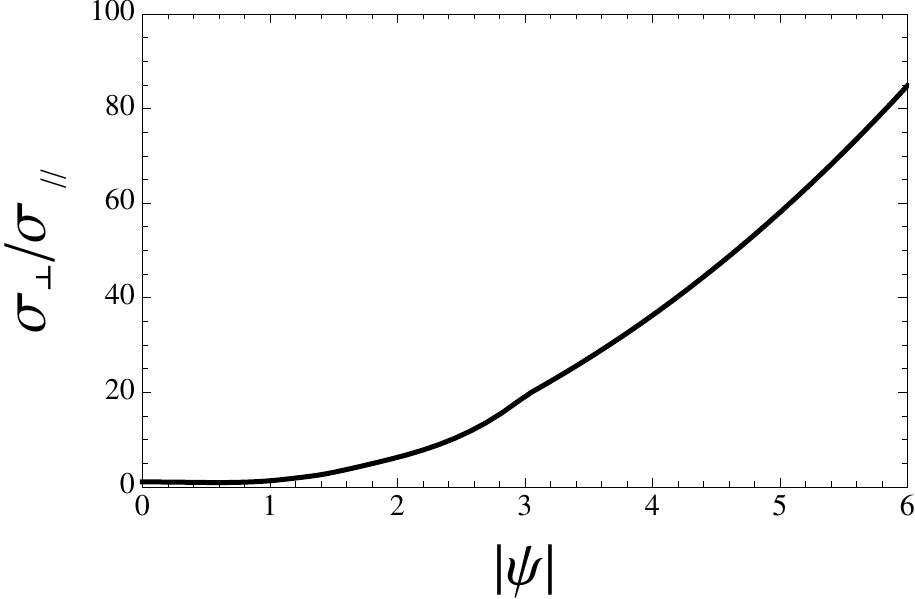}
\caption{Bulk conductivity anisotropy in the B-SDW state, as a function of the amplitude $|\psi|$ of the SDW order parameter.}
  \label{fig:cond-anisotropy}
\end{figure}

We have calculated this conductivity at zero temperature using the
Hartree-Fock quasiparticle Hamiltonian.  From the Boltzmann
equation within relaxation time approximation, one has
\be \sigma_{\mu \nu} = \sum_n
e^2 \tau \int \frac{d^3 k}{8 \pi^3}[ - f'(\epsilon_n ) ]v_{n,\mu} (k) v_{n,\nu} (k) ,
\ee 
where $\tau$ is a constant
relaxation time, $f$ is the Fermi distribution $f(\epsilon)= 1/ (e^{\beta
  (\epsilon-\mu)}+1)$ and $v_{n,\mu} = \partial \epsilon_n(k) / \partial
k_{\mu}$, where $n$ is a band index. We have a total of 8 bands (2 e$_g$ orbitals
$\times$ 4 magnetic sublattices = 8), and the band energies and
velocities must be found numerically.  Using $k \cdot p$ perturbation
theory,\cite{kittel1986introduction} one has:
\begin{equation}
  \label{eq:7}
  v_{n,\mu} (k) = \langle \psi_{nk} | \frac{\partial {\sf H} (k)}{\partial
    k_\mu}| \psi_{nk} \rangle,
\end{equation}
where ${\sf H}(k)$ is the 8$\times$8 matrix Bloch Hamiltonian.  From
the above formulae, we calculated the conductivity $\sigma_\parallel$
parallel to the [111] axis and $\sigma_\perp$ normal to it.  The ratio
is plotted in Fig.~\ref{fig:cond-anisotropy} for the B-SDW state.  As
expected, a large anisotropy is observed once a significant magnetic
order develops.

Note that the same result would be expected to obtain for a thick
film, where the behavior is predominantly bulk-like.  In this case,
the measureable quantity is the effective two-dimensional conductivity
tensor for the plane of the layer, which is usually an [001] plane.
By symmetry, we expect the principle axes of the 2d conductivity to be
the [11] and [1-1] directions, with different conductivities along
each in the SDW state.  Note that in practice this is complicated by
the effects of orthorhombicity, which already should induce transport
anisotropy even in the metallic state.  However, we expect that this
intrinsic anisotropy is probably mild, and that a pronounced effect
due to SDW ordering should be observable below $T_{\rm MIT}$.

\begin{figure}
  \centering
\includegraphics[width=3.in]{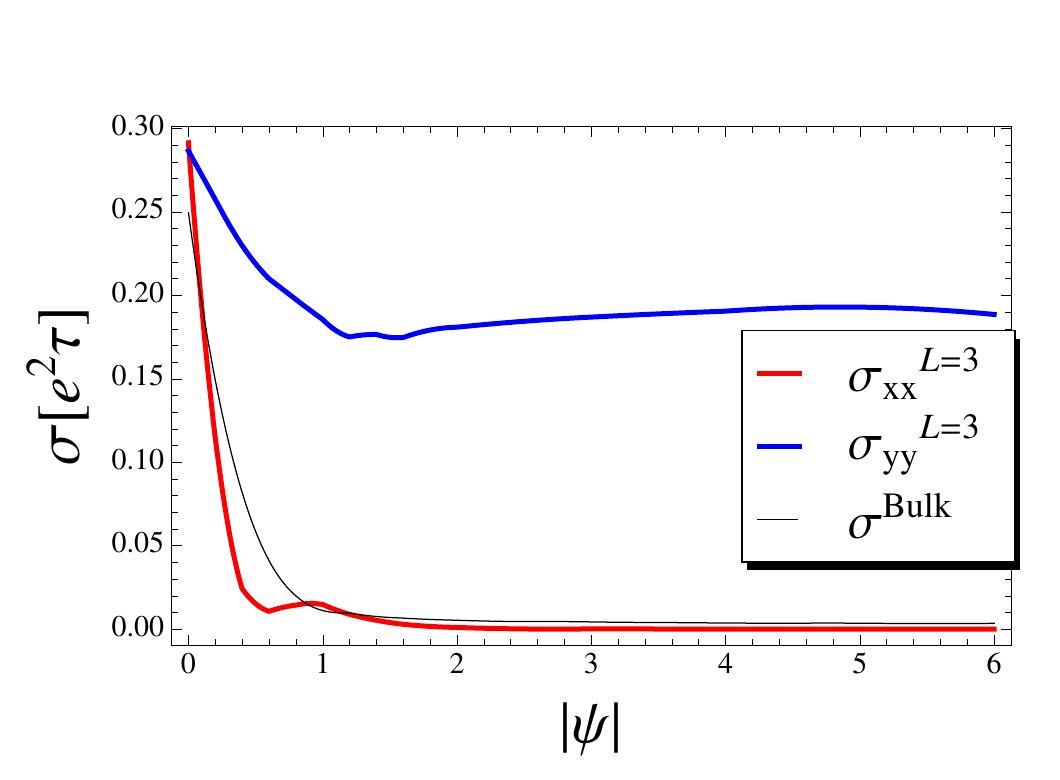}
\caption{Electrical conductivity for $L=3$ and bulk with
  fixed $\tilde t = 1, \tilde t' = 0.15, \theta= \pi/4$ and $\Phi =
  0$. The isotropic conductivity in the bulk case (black line),
  $\sigma^{\text{Bulk}}$, decreases to zero with as a gap in the DOS
  develops with increasing SDW order.   For the three layer case,
  $L=3$, the conductivity shows a large anisotropy in the $x-y$ plane
  once the SDW develops.}
  \label{fig:sigmaL3}
\end{figure}

For thin films, confinement effects may contribute to or modify the
anisotropy.  For instance, in the three layer case, we observed a
change in the nesting wavevector to ${\bf Q}_{L=3}= 2\pi(1/2,0,0)$.
In this state, the anisotropy axes imposed by the SDW are different.
In particular, an``up-up-down-down" magnetic configuration along the
$\hat{x}$ axis is stabilized, so that the spin polarized electrons are
free to hop along ${\hat{y}}$ direction.  Hence, in this case the low
and high conductivity axes are the $[10]$ and $[01]$ axes,
respectively. This is shown in Fig.\ref{fig:sigmaL3}, in which the
magnitude of SDW, $| \psi |$, is varied while fixing $\tilde t = 1,
\tilde t' =0.15, \theta =\pi/4$ and $\Phi =0$.  Indeed, in this case
the anistropic behavior is even more pronounced, for in the model the
``hard'' axis conductivity $\sigma_{xx}$ actually vanishes at $T=0$ in
the limit of large SDW gap, while $\sigma_{yy}$ saturates to a
constant for arbitrarily large $|\psi|$, because the spin polarized
electrons are free to hop along $\hat{y}$ direction. In this case, 
the formation of the SDW opens the Fermi surface.
  
\subsubsection{Optical conductivity }

Optical conductivity is an other important probe of electronic
structure. For LaNiO$_3$, which is metallic at all temperature,
experiment shows a reduced Drude peak compared to band
theory,\cite{dan:opt-cond-LNO} which may be considered as evidence of
moderately strong correlation. However, apart from this quantitative
renormalization of the low energy Drude part, the theoretical optical
conductivity obtained from the simple two e$_g$ band model reproduces
experiment fairly well up to $\omega \approx
2eV$.\cite{dan:opt-cond-LNO} Applying the same analysis to the
magnetically ordered phases in our bulk phase diagram,
Fig.\ref{fig:mit-pd}, we obtained strikingly different results as a
consequence of SDW formation.   

The calculations are made using standard linear response theory within
the Hartree-Fock variational Hamiltonian.   From the Kubo formula, the real part of
optical conductivity $\sigma_{\alpha \beta} ( \Omega, {\bf k}) $ is
related to the imaginary part of current-current correlation
$\Pi_{\alpha \beta} ( \Omega, {\bf k})$: \cite{mahan2000many}
\be
\sigma_{\alpha \beta} = \frac{i}{\omega} \int \frac{d^3 k}{(2 \pi)^3} \Pi_{\alpha \beta} (\Omega,{\bf k}) + \frac{n_0 e^2}{m} \delta{\alpha \beta}
\ee
with wave vector ${\bf k}$, frequency $\Omega$, average density $n_0$
and electron mass $m$.  The current-current correlation function with imaginary frequency $i \Omega_l$ is defined as
\bw
\bea
\Pi_{\alpha \beta} ( i \Omega_l , {\bf k}) =  \frac{2}{vol} \sum_{abcd}  j_{\alpha}^{ab} ({\bf k}) j_\beta^{cd} ({\bf k})  \frac{1}{\beta} \sum_n G_{ad} ( i \omega_n + i \Omega_l,{\bf k}) G_{cb} ( i \omega_n, {\bf k}).
\label{eq:current-corr}
\eea
At zero temperature, this can be calculated from the spectral
representation (see Appendix.\ref{app:opt-cond}),
\bea
{\rm Im}[ \Pi_{\alpha \beta} ( \Omega , {\bf k}) ] = \sum_{m m'} \phi_m^{a*}  \phi_{m'}^{b}  \phi_{m'}^{c*}  \phi_m^{d} \int\frac{d \omega}{\pi} A_m (\omega) A_{m'}  (\omega + \Omega) ( n_F (\omega) - n_F (\omega+\Omega) ),
\eea
\ew
where $A_m (\omega) = \gamma / [ (\omega- E_m+ \mu N)^2 + \gamma^2]$,
with $\gamma$ a small scattering rate (imaginary part of the first order self-energy
correction  ${ \rm Im} [ \Sigma (\omega_n  )] = - i \gamma sgn
(\omega_n)$) added by hand, and  $\phi_m^a ({\bf k})$ is the $a$ component of $m$th
eigenstate. $n_F (\omega) = 1/ (e^{\beta \omega} +1 )$ is Fermi
distribution.  

\begin{figure}
  \centering
\includegraphics[width=7in]{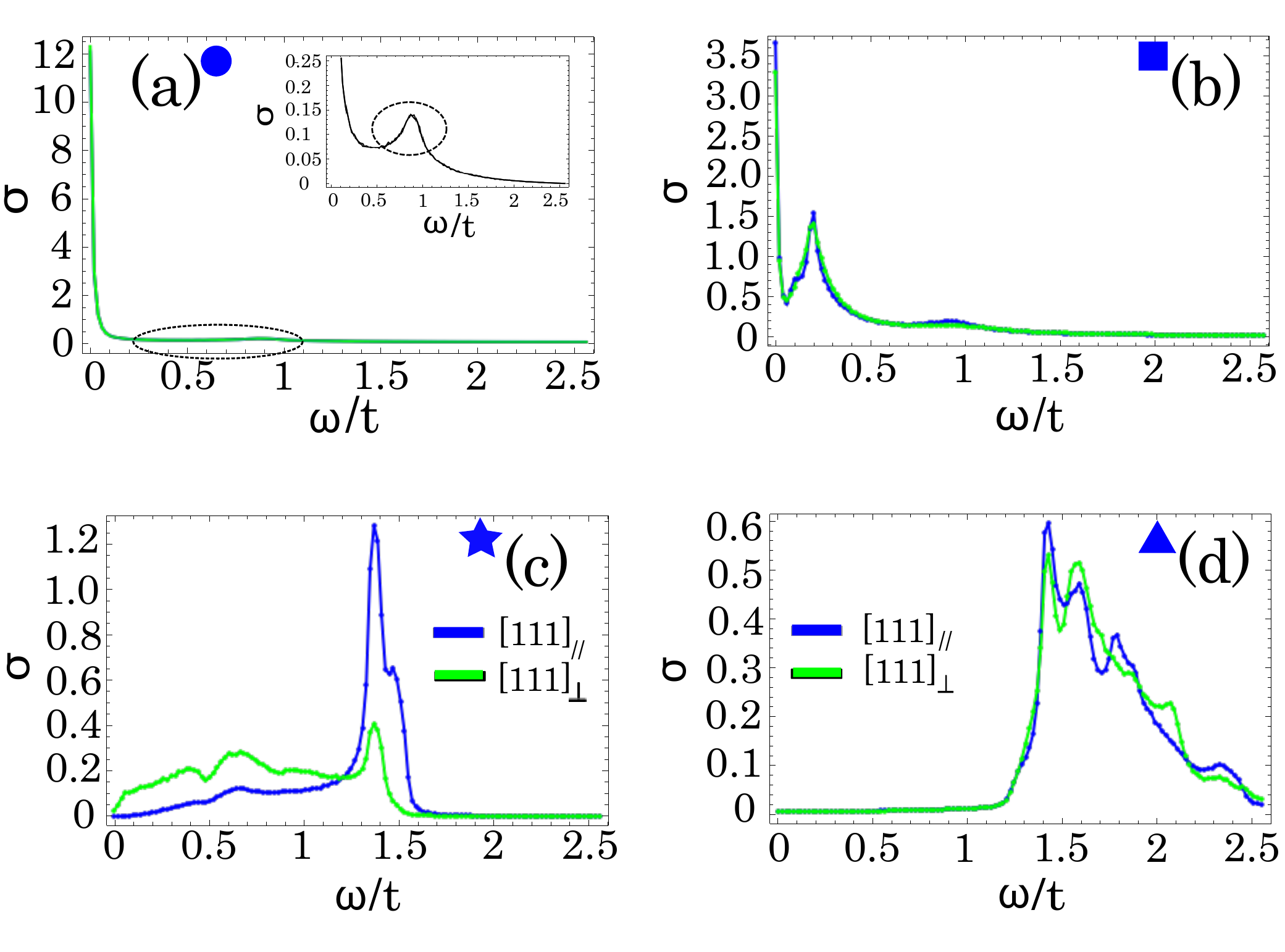}
\caption{The real part of the optical conductivity, $ \sigma
  (\omega)$, for each phase (colored star) in the bulk phase diagram
  (see Fig.\ref{fig:mit-pd}). In (a), the paramagnetic metallic phase
  shows a large Drude peak and a small hump (see inset plot). The hump
  is related to a region of large DOS for interband transitions (see
  Ref.\onlinecite{dan:opt-cond-LNO}).  In (b), the metallic SDW phase has a
  reduced but non-zero Drude peak and a small second peak due to the SDW gap. 
  Plot (c) shows the case of the semi-metallic B-SDW phase, for which a linear increase
  of $\sigma(\omega)$ for small frequency $\omega$ is found, related
  to the linear dispersion near the Fermi level. It also shows strong
  anisotropy between the conductivity  $\sigma_{ [111]_\parallel}$
  (along the [111]
  direction) (blue line) and $\sigma_{ [111]_\perp}$ (perpendicular to
  [111]) (green line).  In plot (d), a large gap is visible in the S-SDW phase. }
  \label{fig:sigma}
\end{figure}

Fig.\ref{fig:sigma} shows the optical conductivity calculated in this
way for each of the different phases (taken at the spots marked by
symbols in the phase diagram in Fig.\ref{fig:mit-pd}).  The
above-mentioned comparison of theory and experiment for the
paramagnetic metallic state is shown in panel (a), taken from
Ref.\onlinecite{dan:opt-cond-LNO}.  The development of SDW order strongly
suppresses the Drude peak, as expected, which can already be seen in
the metallic SDW state when the density of states at the Fermi energy
is still non-zero (but small), Fig.\ref{fig:sigma} (b).
Interestingly, a small peak appears instead at $\omega/t \sim 0.3$. 
This peak arises from a transfer of spectral weight from low frequency
to above the SDW gap.  
Panel (c) shows $\sigma(\omega)$ for the
B-SDW state, which has a semi-metallic band structure.  One observes a
linear increase of ${\rm Re}[\sigma (\omega)]$ for small frequency
$\omega$, which is similar to the behavior expected from the Dirac
points in graphene, and indeed arises from the honeycomb [111] bilayer
structure of the spin-polarized regions, as discussed in
Sec.~\ref{sec:semi-metallic-b}.  Here we have plotted the
powder-average conductivity, since the full tensor is anisotropic as
discussed above.  This calculations has neglected orthorhombicity,
which would introduce a gap at low energy and thereby interupt at
least part of the linear region. However, a linear increase of $\sigma
(\omega)$ at low frequency was indeed seed in bulk experiments on
NdNiO$_3$ below the transition temperature.\cite{katsufuji1995optical}
Finally in Fig.\ref{fig:sigma}(d), we plot the optical conductivity
for for large Hund's coupling $J$, in the S-SDW where strong charge
order is present.  A large gap opens in the spectrum, resulting in
zero ${\rm Re}[\sigma (\omega)]$ up to $\omega/t \approx 1$.

\bw

\ew


\subsection{Summary}
\label{sec:summary}

We have presented a theoretical analysis of the metal-insulator
transition in the nickelates from a minimal two-band model and
Hartree-Fock theory, which we argued is appropriate for the itinerant
limit of weak to intermediate correlation.  This picture of the
metal-insulator transition can be tested in various ways, as suggested
above, and appears to us to be the most consistent one for the materials
NdNiO$_3$ and PrNiO$_3$, located close to the zero temperature MIT phase
boundary.  For the more insulating nickelates, a different type of
theory is required, involving stronger correlation and possibly an
important role for electron-lattice coupling.  Both further theoretical
work in clarifying the mechanism for the MIT transition in those
materials, and experimental work which can test the itinerant picture
(such as measurement of transport anisotropy), would be very
desireable.  Finally, we have shown that quantum confinement alone
cannot explain a Mott insulating in ultrathin LaNiO$_3$ films, and
suggested a physical mechanism by which the observed insulating state
might obtain.  It will be interesting to pursue this question further in
the future.

\acknowledgements

We are grateful to Susanne Stemmer, Jim Allen, Dan Ouellette, and Junwoo
Son for discussions and experimental inspiration.  This work was
supported by the NSF through grants PHY05-51164 and DMR-0804564, and
the Army Research Office through MURI grant No. W911-NF-09-1-0398.

\newpage

\appendix

\section{Dynamical Spin Susceptibility for Free Electrons ${\boldsymbol \chi_0 (\Omega, k)}$}
\label{app:chi_0}

In this section, we derive the dynamical spin susceptibility for free electrons both for bulk and finite layers.
In general, dynamical spin susceptibility for Matsubara frequency $i \Omega_n$, and wave vector $k$ can be represented as following 
\bw
\bea
\chi_0 ( i \Omega_n   , {k} )
 &=&  \langle S^z_{\bf k} S^z_{\bf -k}  \rangle  \label{eq:chi0-def} \\
 &=&  \frac{1}{N} \langle \sum_{{\bf r r' }} S^z_{\bf r} S^z_{\bf r'} e^{ i {\bf k (r-r')} }\rangle   \\
 &=& \frac{1}{N}  \langle \sum_{\bf  r r'} \sum_{\alpha \beta,\alpha' \beta'}  \frac{1}{4} c^{\dagger}_{\bf r \alpha} c_{ \bf r \beta} 
c^{\dagger}_{\bf r' \alpha'} c_{ \bf r' \beta'} \sigma^z_{\alpha \beta} \sigma^z_{\alpha' \beta'} e^{i {\bf k (r-r')}} \rangle \\
&=& \frac{1}{N} \langle \sum_{ \{ \bf q_i \} } \sum_{\bf r r'} \sum_{\alpha \alpha'} \frac{1}{4 N^2} c^{\dagger}_{\bf q_1 \alpha} c_{\bf q_2 \alpha} c^{\dagger}_{\bf q_3 \alpha'} c_{\bf q_4 \alpha'} (-1)^{\alpha+\alpha'}
 e^{i (\bf q_1 - q_2 + k ) r} e^{ i (\bf q_3 - q_4 - k) r' } \rangle \\ 
 &=& \frac{1}{N} \langle \sum_{ \{\bf q_i \} } \sum_{\alpha \alpha'} \frac{1}{4 N^2} N \delta ({\bf q_1 - q_2 + k}) 
 N \delta({\bf q_3 - q_4 - k}) 
 c^{\dagger}_{\bf q_1 \alpha} c_{\bf q_2 \alpha} c^{\dagger}_{\bf q_3 \alpha'} c_{\bf q_4 \alpha'} (-1)^{\alpha + \alpha'} \rangle \\
 &=& \frac{1}{4N}  \langle \sum_{\bf q_1 q_3} \sum_{\alpha \alpha'} 
 c^{\dagger}_{\bf q_1 \alpha} c_{\bf k+q_1  \alpha} c^{\dagger}_{\bf q_3 \alpha'} c_{\bf -k +q_3  \alpha'} (-1)^{\alpha + \alpha'} \rangle \\
 &=&  \frac{1}{2} \int \frac{d^3q}{ (2 \pi)^3}  \frac{1}{\beta} \sum_{\omega_n} \text{Tr} [
 G_0 (i \omega_n,  q) G_0( i(\omega_n + \Omega_n) , q+k) ]
 \label{eq:chi-detail}
\eea
\ew
First of all, ${\bf k (q_i) }$ is four dimensional vector which includes both Matsubara frequency $i \Omega_n (i \omega_n)$ and the wave vector $k ( q_i)$ in three spatial dimension. In the same way, ${\bf r }$ includes both imaginary time $\tau$ and spatial direction $r$. $\alpha, \beta, \alpha'$ and $ \beta'$ are for spin $\uparrow, \downarrow$ and $\sigma$ represents Pauli matrix and ${\bf S}_{\bf r}  =  \sum_{\bf r \alpha \beta} c^{\dagger}_{\bf r \alpha} \frac{ {\boldsymbol \sigma}_{\alpha \beta} }{2} c_{\bf r \beta} $ (with ignoring orbital indices for simplicity). Fourier transform ${\bf S}_{\bf k} = \frac{1}{\sqrt{N}}  \sum_{\bf r} {\bf S}_{\bf r} e^{ i {\bf k \cdot  r} }$ and $c^\dagger_{\bf r \alpha} = \frac{1}{\sqrt{N}}c^\dagger_{\bf q \alpha} e^{i {\bf q \cdot r}}$, free electron Green's function  $G_0 ( i \omega_n , q) = \langle c^\dagger_{\bf q}  c_{\bf q} \rangle = ( i \omega_n - E_q ) ^{-1}$. From the last equation of Eq.~\eqref{eq:chi-detail}, we sum all the Matsubara frequenies using the following trick 
\bw
\bea
\frac{1}{\beta} \sum_{\omega_n} \frac{1}{i \omega_n -x } \frac{1}{ i ( \omega_n + \Omega_n) -x'} 
= \frac{-1}{i \Omega_n +x -x'} ( n_F(x) - n_F (x') )
\eea
\ew
where Fermion distribution is defined as $n_F(x) = 1/(e^{\beta x} +1 )$. 
For simplicity, we represent doubly degenerate e$_g$ orbitals tight-binding model using Pauli matrices $\sigma$, $H_{tb} ( k ) = \epsilon_0 (k) {\bf 1} + \epsilon (k) \cdot  {\boldsymbol \sigma}$. Then finally analytic continuation leads 
\bw
\bea
 \chi_0 ( \Omega, k) 
= \frac{1}{2} \int \frac{d^3 q}{(2 \pi)^3} \sum_{a,b \in \pm} \frac{-1}{\Omega + x_a (q) - x_b ( k+q)}
( n_F (x_a (q) ) - n_F( x_b (k+q)) ) \frac{1}{2} (1+ a b  \frac{\epsilon(q) \cdot \epsilon (k+q) }{ | \epsilon(q) | | \epsilon (k+q) |}  ) 
\eea
\ew
where $x_{\pm} (k)  = \epsilon_0 (k)  \pm | \epsilon (k) | - \mu$.

For finite layers (along $\hat{z}$ direction), it has discretized $k_z = \pi l_z  / (N_z +1)$ where $l_z \in \{ {1, 2, 3 \cdots N_z} \}$ and  $N_z$ is the number of layers. 
\be
c^{\dagger}_{r \alpha} = \sqrt{\frac{2}{N_{\perp}N_z}} \sum_k c^{\dagger}_{k \alpha} e^{i k_{\perp} r_{\perp} } \sin {k_z r_z}
\label{eq:c_fin}
\ee
Here, $N_{\perp}$ is the number of sites on its perpendicular $x-y$ plane. The spin susceptibility Eq.~\eqref{eq:chi0-def} is represented 
\bw
\bea 
\label{eq:app-chi0-fin-1}
\langle S^z_{\bf k} S^z_{\bf -k} \rangle 
&=& \frac{1}{N_\perp N_z} \langle \sum_{ \{  q^{\perp}_i \} }  \sum_{ \{ q^{z}_i \} } \sum_{\alpha \alpha'} 
\frac{2^2}{ (N_\perp N_z)^2} \frac{ (-1)^{\alpha + \alpha'} }{4}
c^{\dagger}_{ q _1 \alpha} c_{ q _2 \alpha }  c^{\dagger}_{q _3 \alpha' }  c_{q _4 \alpha'} 
e^{ i (q^ \perp_{1} - q^{\perp}_2 + k_{\perp} ) r_\perp  }
e^{ i (q^{\perp}_3 - q^{\perp}_4 + k_{\perp} ) r'_\perp  } \nonumber \\
 & & \hspace{5cm} \sin q^z_1 r_z \sin q^z_2 r_z \sin q^z_3 r'_z  \sin q^z_4 r'_z  e^{ i k_z (r_z -r'_z) }  \rangle \\
 \label{eq:app-chi0-fin-2}
&=& \frac{1}{N_\perp N_z} \langle \sum_{ \{q^\perp_i \} } \sum_{ \{ q^z_i \} } \sum_{\alpha \alpha'}
  \frac{2^2}{ ( N_\perp N_z )^2 }  \frac{ (-1)^{\alpha +\alpha'} }{4}  
 c^{\dagger}_{q_1 \alpha} c_{q_2 \alpha} c^{\dagger}_{q_3 \alpha'} c_{q_4 \alpha '} 
 N_\perp \delta ( q^\perp_1 - q^\perp_2 + k_\perp )  N_\perp \delta ( q^\perp_3 - q^\perp_4  - k_\perp ) \nonumber \\
& & \hspace{3cm} \frac{1}{2^4} N_z ( \sum_{ab \in \pm } \delta ( a q^z_1 + b q^z_2 + k_z )  )
 N_z ( \sum_{ab \in \pm } \delta ( a q^z_3 + b q^z_4 - k_z )  ) \rangle \\
 \label{eq:app-chi0-fin-3}
 &=& \frac{1}{N_\perp N_z} \langle \sum_{q^\perp_1 q^\perp_3} \sum_{ \{ q_i^z \} } \sum_{\alpha \alpha'} 
 \frac{(-1)^{\alpha \alpha'} }{2^4} c^\dagger_{q^\perp_1  q^z_1 \alpha } c_{k_\perp+q^\perp_1  q^z_2 \alpha } 
 c^\dagger_{q^\perp_3  q^z_3 \alpha' } c_{-k_\perp+q^\perp_3  q^z_4 \alpha' } \nonumber \\
& &\hspace{5cm}  ( \sum_{ab \in \pm } \delta ( a q^z_1 + b q^z_2 + k_z )  ) ( \sum_{ab \in \pm } \delta ( a q^z_3 + b q^z_4 - k_z )  )  \rangle \\
\label{eq:app-chi0-fin-4}
&=& \frac{1}{2^3 N_z} \sum_{q^z_1 q^z_2}\int \frac{d^2 q_\perp}{ (2\pi)^2} \frac{1}{\beta} \sum_{\omega_n}
Tr [ G_0 ( i \omega_n , q_\perp, q^z_1 ) 
G_0 ( i (\omega_n + \Omega_n)  , (q_\perp + k_\perp ) ,  q^z_2   ) \nonumber \\
& & \hspace{2cm} (  \delta (  q^z_1 +  q^z_2 + k_z )  + \delta (  q^z_1 +  q^z_2 - k_z ) 
- \delta (  q^z_1 -  q^z_2 + k_z ) - \delta (  q^z_1 -  q^z_2 - k_z )  )^2   ] 
\eea 
\ew
From Eq.~\eqref{eq:app-chi0-fin-1} to \ref{eq:app-chi0-fin-3}, we abbreviate Matsubara frequency indices for simple representation. $\delta$ functions in Eq.~\eqref{eq:app-chi0-fin-4} can be rewritten 
$\delta ( q_1^z +q_2^z + k_z) = \delta (l_1 + l_2 + l_z ) (\text{mod} (2 (N_z +1 ) ))$ where $q^z_i  = \pi l_i / ( 2N_z +1)$ and $k_z = \pi l_z / (2N_z +1)$.

\section{Detailed Hartree-Fock Calculation} 
\label{app:detail-HF}
 
Our variational Hamiltonian Eq.~\eqref{eq:tildeH-tb-8} can be diagonalized by writing 
\be
  \label{eq:14}
  c_{na\alpha}(k) = \sum_{A} \phi^{A\alpha}_{na}(k) c_{A\alpha}(k),
\ee 
with $A=1\ldots 8$ indexes the eigenstates for each $k$.  With
appropriate choice of $\phi$ the diagonalized Hamiltonian becomes
\begin{equation}
  \label{eq:15}
  H_{\rm var} = \sum'_k \sum_{A\alpha} \epsilon_{A}(k)
  c_{A\alpha}^\dagger(k) c_{A\alpha}^{\vphantom\dagger}(k).
\end{equation}
Here we have used that $\epsilon_A(k)$ are independent of $\alpha$.
This can be seen since the transformation $c_{na\alpha}(k) \rightarrow
(-1)^n c_{na\alpha}(k)$ maps $\alpha\rightarrow -\alpha$.  Hence
\begin{equation}
  \label{eq:16}
  \phi^{A-}_{na}(k) = (-1)^n \phi^{A+}_{na}(k),
\end{equation}
and the energies are independent of $\alpha$. Now we take the expectation values of each term. Ground state $ | \Psi_0 \rangle$ is nothing but occupying all the quasiparticles states below the Fermi energy. First, we consider the expectation value of $H_{\rm tb}$. 
\bw
\be
  \label{eq:18}
 \langle  H_{\rm tb} \rangle = \sum'_k \sum_n \sum_{ab} \sum_A {\mathcal H}_{ab}(k+n Q)
  \left( \phi^{A\alpha}_{n a}(k) \right)^* \phi^{A\alpha}_{nb}(k) n_F(\epsilon_A(k)).
\ee
where $n_F (\epsilon)$ is the Fermi function. Next, consider the expectation value of on-site Coulomb interaction $H_U = U \sum_i n_i^2$.  
\bea
  \label{eq:19}
  \sum_i n_i^2 & = & \sum_i c_{ia\alpha}^\dagger
  c_{ia\alpha}^{\vphantom\dagger} c_{ib\beta}^\dagger
  c_{ib\beta}^{\vphantom\dagger} \nonumber \\
  & = & \frac{1}{N} \sum_{k_1 k_2 k_3 k_4} c_{a\alpha}^\dagger (k_1)
  c_{a\alpha}^{\vphantom\dagger} (k_2)  c_{b\beta}^\dagger (k_3)
  c_{b\beta}^{\vphantom\dagger} (k_4)  \delta_{k_1+k_3,k_2+k_4}
  \nonumber \\
  & = & \frac{1}{N} \sum'_{\{ k_i \}} \sum_{\{ n_i \}} c_{n_1 a\alpha}^\dagger (k_1)
  c_{n_2 a\alpha}^{\vphantom\dagger} (k_2)  c_{n_3 b\beta}^\dagger (k_3)
  c_{n_4 b\beta}^{\vphantom\dagger} (k_4)  \delta_{k_1+k_3,k_2+k_4}
  \delta_{n_1+n_3,n_2+n_4} (\textrm{mod 4}) \nonumber \\
& = & \frac{1}{N} \sum'_{\{ k_i \}} \sum_{\{ n_i \}} \sum_{ABCD} \left(
  \phi^{A\alpha}_{n_1 a}(k_1)\right)^* \phi^{B\alpha}_{n_2 a}(k_2) \left(
  \phi^{C\beta}_{n_3 b}(k_3)\right)^* \phi^{D\beta}_{n_4 b}(k_4)
\nonumber \\
& & \times
c_{A\alpha}^\dagger (k_1) c_{B\alpha}^{\vphantom\dagger}(k_2)
c_{C\beta}^\dagger (k_3)c_{D\beta}^{\vphantom\dagger}(k_4) 
\delta_{k_1+k_3,k_2+k_4}
  \delta_{n_1+n_3,n_2+n_4} (\textrm{mod 4}).
\eea
\ew
Now we take the expectation value. There are both Hartree and Fock terms.
\bw
\bea
  \label{eq:20}
  \left\langle \sum_i n_i^2\right\rangle
  & = & \frac{1}{N} \sum'_{k_1,k_3}  \sum_{\{ n_i \}} \sum_{A,C} \sum_{ab} \sum_{\alpha\beta} \Big[ \left(
    \phi^{A\alpha}_{n_1 a}(k_1)\right)^* \phi^{A\alpha}_{n_2 a}(k_1) \left(
    \phi^{C\beta}_{n_3 b}(k_3)\right)^* \phi^{C\beta}_{n_4 b}(k_3)
  \nonumber \\
 && -\left(\phi^{A\alpha}_{n_1 a}(k_1)\right)^* \phi^{C\alpha}_{n_2 a}(k_3) \left(
    \phi^{C\alpha}_{n_3 b}(k_3)\right)^* \phi^{A\alpha}_{n_4
    b}(k_1)\delta_{\alpha\beta} \Big]n_F(\epsilon_A(k_1)) n_F(\epsilon_C(k_3)) \delta_{n_1+n_3,n_2+n_4} (\textrm{mod 4})
\eea
\ew
Using Eq.~\eqref{eq:16}, Eq.~\eqref{eq:20} can be simplified after summing the spin indices $\alpha, \beta$
\bw
\bea 
\label{eq:sim-U}
\left\langle \sum_i n_i^2 \right\rangle &=& \frac{1}{N} \sum_{k_1 k_3}  \sum_{ \{ n_i \} }\sum_{AC} \sum_{ab}  
[  2( 1+(-1)^{n_1+n_2} ) 
    (\phi^{A +}_{n_1 a}(k_1) )^* \phi^{A +}_{n_2 a}(k_1) (
    \phi^{C +}_{n_3 b}(k_3) )^* \phi^{C +}_{n_4 b}(k_3)  \nonumber \\
&&   - 2  \left(\phi^{A +}_{n_1 a}(k_1)\right)^* \phi^{C +}_{n_2 a}(k_3) \left(
    \phi^{C +}_{n_3 b}(k_3)\right)^* \phi^{A +}_{n_4 b}(k_1) ] n_F(\epsilon_A(k_1)) n_F(\epsilon_C(k_3)) \delta_{n_1+n_3,n_2+n_4} (\textrm{mod 4})
\eea
\ew
In the same way, the expectation value of Hund's coupling is represented by 
\bw
\bea
\label{eq:hunds}
\left\langle  \sum_i {\bf S}_i^2 \right\rangle  &=&
\frac{1}{4N} \sum'_{k_i} \sum_{\{ n_i \} } \sum_{ABCD} [ (\phi_{n_1 a}^{A \alpha} (k_1) )^*  {\boldsymbol \sigma}_{\alpha \beta} \phi_{n_2 a}^{B \beta} (k_2) ] 
\cdot [ (\phi_{n_3 b}^{C \alpha'} (k_3) )^*  {\boldsymbol \sigma}_{\alpha' \beta'} \phi_{n_4 b}^{D \beta'} (k_4) ]  
n_F ( \epsilon_A (k_1) ) n_F ( \epsilon_C (k_3) )
\nonumber \\
&& [\delta_{AB} \delta_{CD} \delta_{k_1 k_2} \delta_{k_3 k_4} \delta_{\alpha \beta} \delta_{\alpha' \beta'}  
- \delta_{AD} \delta_{BC} \delta_{k_1 k_4} \delta_{k_2 k_3} \delta_{\alpha \beta'} \delta_{\beta \alpha'}  ] 
 \delta_{n_1+n_3, n_2+n_4} (\textrm{mod 4} ) \nonumber \\
 &=&  \frac{1}{4N} \sum'_{k_1 k_3} \sum_{ \{n_i \} } \sum_{AC} \sum_{\alpha \beta \alpha' \beta'}
 [ ( \phi_{n_1 a}^{A \alpha} (k_1) )^*  \phi_{n_2 a}^{A \beta} (k_1) 
  (\phi_{n_3 b}^{C \alpha'} (k_3) )^* \phi_{n_4 b}^{C \beta'} (k_3)  
  ( 2 \delta_{\alpha \beta \alpha' \beta' } -\delta_{\alpha \beta} \delta_{\alpha' \beta'} ) \nonumber \\
&&  -  ( \phi_{n_1 a}^{A \alpha} (k_1) )^*  \phi_{n_2 a}^{C \beta} (k_3) 
  (\phi_{n_3 b}^{C \alpha'} (k_3) )^* \phi_{n_4 b}^{A \beta'} (k_1)  
  ( 2 \delta_{\alpha \beta'} \delta_{\beta \alpha'}  -\delta_{\alpha \beta \alpha' \beta'} ) ]  
 \nonumber \\
 & &  n_F(\epsilon_A(k_1)) n_F(\epsilon_C(k_3)) \delta_{n_1+n_3,n_2+n_4} (\textrm{mod 4}) \nonumber  \\
&=&   \frac{1}{4N} \sum'_{k_1 k_3} \sum_{\{n_i\} } \sum_{AC} \sum_{ab}  
[ 2(1- (-1)^{n_1 + n_2} )  
  (\phi^{A +}_{n_1 a}(k_1) )^* \phi^{A +}_{n_2 a}(k_1) (
    \phi^{C +}_{n_3 b}(k_3) )^* \phi^{C +}_{n_4 b}(k_3)  \nonumber \\
&& - (2 + 4 (-1)^{n_1+n_4} )  
 (\phi^{A +}_{n_1 a}(k_1) )^* \phi^{C +}_{n_2 a}(k_3) (
    \phi^{C +}_{n_3 b}(k_3) )^* \phi^{A +}_{n_4 b}(k_1) ] \nonumber \\
   && n_F(\epsilon_A(k_1)) n_F(\epsilon_C(k_3)) \delta_{n_1+n_3,n_2+n_4} (\textrm{mod 4})
\eea
\ew
Finite layers $N_z$ along $\hat{z}$ direction lead the discretized $k_z = \pi l_z /(L+1)$ where $l = 1,2, 3 \cdots N_z$
\be
c_{i a \alpha} = \sqrt{ \frac{2}{N} } \sum'_{ k_{\perp} k_z}\sum_n c_{ n a \alpha} (k_\perp k_z)  e^{ i k_\perp r_{i \perp}} \sin{k_z r_z}
\ee
Now, the interaction term Eq.~\eqref{eq:20} has modified function of $\delta_{k_1+k_3,k_2+k_4}$ For both Eq.~\eqref{eq:20} and Eq~\eqref{eq:hunds} need $1/4  \delta_{k_\perp^1 + k_\perp^3, k_\perp^2 + k_\perp^4} ( 4 + 2 \delta_{2(k_z^1+k_z^3)} (\text{mod} (2\pi)) + 2 \delta_{2(k_z^1-k_z^3) } (\text{mod} (2\pi))$
The last two $\delta$ functions correspond to 
$ \delta_{(l_1 + l_3) , L+1}$ and $\delta_{l_1,l_3}$.

\section{Orthorhombic GdFeO$_3$ distortion}
\label{app:ortho}
In this section, we discuss the point symmetries of the orthorhombic lattice (GdFeO$_3$ type perovskite) and study how this symmetry operators constraint on-site splitting vectors ${\bf D}_i$ defined in Eq.~\eqref{eq:H_ortho}.
We first the define four basis sites of the orthorhombic lattice (in cubic coordinates):
\begin{eqnarray}
  \label{eq:39}
  {\bm r}_1 & = & (0,0,0), \\
  {\bm r}_2 & = & (1,0,0), \\
  {\bm r}_3 & = & (0,0,1), \\
  {\bm r}_4 & = & (1,0,1).
\end{eqnarray}
The orthorhombic space group has three point group operations (in cubic coordinates) : 
\begin{eqnarray}
  \label{eq:42}
    P_1: ({x,y,z}) & \longrightarrow & (1-y,-x, -z), \\
   P_2: ({x,y,z}) & \longrightarrow & (-1-x,1-y,1+z), \\
   P_3: ({x,y,z}) & \longrightarrow & (-1-x,1-y,-z).
\end{eqnarray}
 One finds that $P_1$
interchanges sites ${\bm r}_1 \leftrightarrow {\bm r}_2$, and ${\bm r}_3
\leftrightarrow {\bm r}_4$, while $P_2$ interchanges sites ${\bm r}_1
\leftrightarrow {\bm r}_3$, and ${\bm r}_2\leftrightarrow {\bm r}_4$.
The inversion $P_3$ leaves the basis unpermuted.
Taking the usual cubic basis of $d_{x^2-y^2}$ and $d_{z^2}$ orbitals,
one then readily finds the transformations of creation/annihilation
operators:
\begin{equation}
  \label{eq:43}
    P_1: \left\{ \begin{array}{ccc}
  {c}_1 & \to &  -\tau^z {c}_2 \\
  {c}_2 & \to & -\tau^z {c}_1 \\
  {c}_3 & \to & -\tau^z {c}_4 \\
  {c}_4 & \to & -\tau^z {
    c}_3 \end{array}\right. \qquad
  P_2: \left\{ \begin{array}{ccc}
  {c}_1 & \to &  {c}_3 \\
  {c}_2 & \to & {c}_4\\
  {c}_3 & \to &  {c}_1 \\
  {c}_4 & \to &  {c}_2 \end{array}\right. \qquad
P_3: {\mathcal I},
\end{equation}
where the last equation indicates that $P_3$ acts as the identity in
both the orbital and sublattice space.
From this we see that $P_3$ places no constraints whatsoever on the
orbital fields.  Invariance under the first and second transformations
then allows all four orbital fields to be determined from one.  One
finds:
\begin{eqnarray}
  \label{eq:44}
  {\bm D}_1 & = & (D^x,D^y,D^z), \\
  {\bm D}_2 & = & (-D^x,-D^y,D^z), \\
  {\bm D}_3 & = & (D^x,D^y,D^z), \\
  {\bm D}_4 & = & (-D^x,-D^y,D^z).
\end{eqnarray}
Thus there are two and not four different orbital fields appearing.
Taking into account the coordinates of these basis sites, we can finally
write a simple form which is basis independent:
\begin{equation}
  \label{eq:45}
  {\bm D}_i = \left( (-1)^{x_i+y_i} D^x,(-1)^{x_i+y_i}D^y,D^z\right).
\end{equation}

\section{Degenerate perturbation theory calculation in the strong coupling limit}
\label{app:DPT}
\subsection{$O(t^2/J_H)$: charge ordering}
\label{subsec:elec-loc}
There are three possible types of hops at second order:
\begin{enumerate}
\item An electron hops from a double
occupied site to an empty site, and then back. This lowers the energy
when occupied sites are adjacent to empty sites, and so results in an
effective repulsion between boson pairs.
\item Both electrons from a
doubly occupied site hop onto the same, previously empty, site. This
results in an effective hopping of the bosons.
\item In the case where neighboring sites are occupied with bosons,
  there can be exchange if the spins of both bosons are not parallel.
\end{enumerate}
The terms in the effective Hamiltonian
corresponding to the above three procedures can be written as
\begin{eqnarray}
  \label{eq:62}
\mathcal{H}_{\rm eff}^{(1)} & = & - \mathcal{P} t_{ij}^{ab}
  c_{ia\alpha}^\dagger c_{jb\alpha}^{\vphantom\dagger} \mathcal{R}
  \mathcal{Q} t_{ji}^{ba} c_{jb\alpha}^\dagger
  c_{ia\alpha}^{\vphantom\dagger} \mathcal{P},  \\
\mathcal{H}_{\rm eff}^{(2)} & = & - 
  \mathcal{P} t_{ij}^{ab} c_{ia\alpha}^\dagger c_{jb\alpha}^{\vphantom\dagger} \mathcal{R}
  \mathcal{Q} t_{ij}^{cd}c_{ic\beta}^\dagger c_{jd\beta}^{\vphantom\dagger}\mathcal{P} , \\
 \mathcal{H}_{\rm eff}^{(3)} &=& -2
  \mathcal{P}  t_{ij}^{ab}c_{ia\alpha}^\dagger c_{jb\alpha}^{\vphantom\dagger} \mathcal{R}  \mathcal{Q}
  t_{ji}^{ba}c_{jb\beta}^\dagger c_{ia\beta}^{\vphantom\dagger} \mathcal{P}.
\end{eqnarray}
All three terms include implied sums over nearest-neighbor sites $i$
and $j$.  Here we have neglected $O(t'^2)$ contributions which are
parametrically small in the limit considered. The factor 2 in
Eq.(23) arises from the fact that electrons can hop first
from site $i$ to $j$ or vice-versa. 

Using the exact form of the hopping matrix in
Eq.~\eqref{eq:tight-binding}, one finds that the second effective
Hamiltonian vanishes, $\mathcal{H}_{\rm eff}^{(2)} =0$.  This can also
be understood from simple orbital considerations: only one of the two
orbitals overlaps along any of the principle directions.  Since both
electrons must be transferred for the pair to transfer, the boson
hopping vanishes. Due to the absence of the pair hopping, the effective Hamiltonian commutes with $N_i$.

\subsection{Magnetic interactions}
\label{subsec:spin-state-boson}
Consider a plaquette on the original cubic lattice, we name the
occupied sites $i$ and $j$ and empty sites $k$ and $j$, such that $i$
and $j$ are next nearest-neighbor on the square plaquette.  We
calculate terms for nearest neighbor spin exchange at different orders one by one as follows.
\begin{enumerate}
\item $O(t^4/J_H^3)$:
  \begin{enumerate}
    \item One electron from each of sites $i$ and $j$ hops to site $k$, and then the two electrons at site $k$ return to $i$ and $j$. There are four distinct time orders in which this process can occur and they contribute equally. The same procedure can also happen to sites $i$, $j$ and $l$. This give the coefficient 8 in front the Hamiltonian below
\bea
  \label{eq:65}
 \mathcal{H}_{\rm eff}^{(4)} & =  - 8 \mathcal{P}
  t_{jk}^{cf} c_{jc\gamma}^\dagger c_{kf\gamma}^{\vphantom\dagger} \mathcal{R}\mathcal{Q}
  t_{ik}^{ae} c_{ia\lambda}^\dagger c_{ke\lambda}^{\vphantom\dagger} \nonumber \\
 &\mathcal{R}\mathcal{Q} t_{kj}^{dc} c_{kd\alpha}^\dagger c_{jc\alpha}^{\vphantom\dagger} \mathcal{R}\mathcal{Q} t_{ki}^{ba} c_{kb\beta}^\dagger c_{ia\beta}^{\vphantom\dagger}\mathcal{P} .
\eea
\item One electron at site $i$ hops to site $k$ and then to $j$, it forms
a singlet state with another electron at site $j$, and then one of two
electrons forming a singlet hops back to site $k$ and then to $i$  (For
brevity, we will not write down the effective Hamiltonian of the other
hopping procedure from now on)
\item One electron from site $i$($j$) hops to site $k$($l$), so now four
  corners of the plaquette are all occupied with single electrons, then
  the electron at site $k$($l$) hops back to $j$($i$)
\item One electron at site $i$ hops to site $k$ and then to $j$, it
  forms a singlet state with another electron at site $j$, and then one
  of two electrons forming a singlet hops to site $l$ and then to $i$. 
\end{enumerate}
Combining the four terms, we have
\bea
  \label{eq:67}
J_1^{(1)} &\equiv& J_{i,i\pm\hat{\mu}\pm \hat{\nu}} ^{(1)} \nonumber \\
 &=& -\frac{t^4}{J_H^3}\frac{1}{(1-4\alpha)^2}\big[\frac{8(5+4\alpha)}{(1-4\alpha)(5-4\alpha)} \nonumber \\
&&-\frac{5}{5+4\alpha}-\frac{1}{1-4\alpha}\big].
\eea

\item $O(t^2t'/J_H^2)$:
\begin{enumerate}
\item One electron at site $i$ hops to $j$ via next nearest-neighboring
  hopping, it forms a singlet state with another electron at site $j$,
  then one of two electrons forming a singlet hops back to site $k$ and
  then to $i$
\item One electron at site $i$ hops to $k$, another electron at site $j$
  hops to site $i$ via next nearest-neighbor hopping, then the electron
  at site $k$ hops to $j$. 
\end{enumerate}
Together these two terms give
\be
  \label{eq:68}
 J_1^{(2)} =  \frac{t^2t'}{J_H^2}\frac{1}{1-4\alpha}\left[\frac{10}{5+4\alpha}+\frac{5}{1-4\alpha}\right].
\ee

\item $O(t'^2/J_H)$:

  One electron at site $i$ hops to an fcc nearest-neighbor $j$, forming a
  singlet state with another electron at site $j$, then one of the two
  electrons forming the singlet hops back to site $i$. We obtain
\be
  \label{eq:69}
 J_1^{(3)} =  \frac{t'^2}{J_H}\frac{5}{5+4\alpha}.
\ee

\end{enumerate}

The spin exchange coupling between nearest neighbor is then $J_1 = J_1^{(1)}+J_1^{(2)}+J_1^{(3)}$.

For second nearest-neighbor spin exchange, consider three sites $i$,$k$ and $j$ along the same cubic axis, where $i$($j$) and $k$ are nearest neighbors of the original cubic lattice. Sites $i$ and $j$ then correspond to the second nearest-neighbor sites. Then there are two possible ways of the hopping procedure, which is of identical hopping order to the first two cases of $O(t^4/J_H^3)$ terms.

\section{Optical conductivity}
\label{app:opt-cond}
 The current-current correlation function with imaginary time is defined as
\bw
\be 
\Pi_{\alpha \beta} (\tau, {\bf k}) = - \frac{1}{vol} \langle T_{\tau- \tau'} j_\alpha^\dagger (\tau, {\bf k}) j_{\beta} (\tau', {\bf k})\rangle \nonumber \\
= \frac{2}{vol} \sum_{abcd} j_\alpha^{ab} ({\bf k} ) j_\beta^{cd} ({\bf k} ) G_{ad} (\tau' -\tau ,{\bf k} ) G_{cb} (\tau- \tau' ,{\bf k})
\ee
\ew
where $j_{\alpha}^{ab} = \partial H^{ab} / \partial k_\alpha$, $G_{ad} ( \tau'-\tau,{\bf k} )$ is a retarded Green's function with imaginary time $\tau'-\tau$ and wave vector ${\bf k}$ and a prefactor 2 for spin sums.
Fourier transform with Matsubara frequency $i \omega_n$ leads
\bw
\bea
\Pi_{\alpha \beta} ( i \Omega_l , {\bf k}) =  \frac{2}{vol} \sum_{abcd}  j_{\alpha}^{ab} ({\bf k}) j_\beta^{cd} ({\bf k})  \frac{1}{\beta} \sum_n G_{ad} ( i \omega_n + i \Omega_l,{\bf k}) G_{cb} ( i \omega_n, {\bf k})
\label{eq:current-corr}
\eea
\ew
Green's function $G_{ad} (i \omega_n,{\bf k})$,
\bea
G_{ad} (i \omega_n,{\bf k}) &=& \int  d\tau \Theta (\tau) G_{ad} (\tau, {\bf k}) e^{ i \omega_n \tau}  \nonumber \\
&=&\sum_{m m'} \frac{e^{-\beta \epsilon_m } + e^{-\beta \epsilon_n}}{ i \omega_n + \epsilon_m' -\epsilon_m } \langle m' | c_d |m  \rangle \langle m | c_a^\dagger | m' \rangle \nonumber \\
&=& \sum_m \frac{\phi_m^{a *} ({\bf k}) \phi_m^d ({\bf k})} 
 { i \omega_n + i \gamma  sgn(\omega_n) - ( E_m -\mu N ) }
\label{eq:greens}
 \eea
The last term in Eq.~\eqref{eq:greens} is for zero temperature with the imaginary part of the first order self-energy correction  ${ \rm Im} [ \Sigma (\omega_n  )] = - i \gamma sgn (\omega_n)$, the $a$ component of $m$ eigenstate $\phi_m^a ({\bf k})$. 
By substituting Eq~\eqref{eq:greens} to Eq.~\eqref{eq:current-corr} and using analytic continuation $ i \Omega_l \rightarrow \Omega + i \eta$, the imaginary part of the current-current correlation function is represented as
\bw
\bea
{\rm Im}[ \Pi_{\alpha \beta} ( \Omega , {\bf k}) ] = \sum_{m m'} \phi_m^{a*}  \phi_{m'}^{b}  \phi_{m'}^{c*}  \phi_m^{d} \int\frac{d \omega}{\pi} A_m (\omega) A_{m'}  (\omega + \Omega) ( n_F (\omega) - n_F (\omega+\Omega) )
\eea
\ew
where $A_m (\omega) = \gamma / [ (\omega- E_m+ \mu N)^2 + \gamma^2]$ and Fermi distribution $n_F (\omega) = 1/ (e^{\beta \omega} +1 )$.


\bibliography{HF-MITn}

\end{document}